\documentclass[twocolumn,amsmath,amssymb,floatfix,superscriptaddress,prb]{revtex4-2}

\usepackage{dcolumn}
\usepackage{bm,graphicx,url,epsf,color}
\usepackage{amssymb}
\usepackage{amsthm}
\usepackage[colorlinks=true,linkcolor=blue,citecolor=blue]{hyperref}
\usepackage{comment}
\usepackage{physics}

\newcommand{\br}{\bm r}
\newcommand{\bk}{\bm k}
\newcommand{\bq}{\bm q}
\newcommand{\eqn}[1]{(\ref{#1})}

\newcommand{\bR}{\bm R}
\newcommand{\bp}{\bm p}
\newcommand{\mr}{moir\'e~}

\newcommand{\mom}{supermoir\'e~}
\newcommand{\bKM}{{\bf K}_\text{M}}
\newcommand{\bGM}{{\bf \Gamma}_\text{M}}

\begin{document}
\title{Supermoir\'e low-energy effective theory of twisted trilayer graphene}

\author{Yuncheng Mao}
\affiliation{Universit\'e Paris Cit\'e, CNRS,  Laboratoire  Mat\'eriaux  et  Ph\'enom\`enes  Quantiques, 75013  Paris,  France}
\author{Daniele Guerci}
\affiliation{Center for Computational Quantum Physics, Flatiron Institute, New York, New York 10010, USA}
\author{Christophe Mora}
\affiliation{Universit\'e Paris Cit\'e, CNRS,  Laboratoire  Mat\'eriaux  et  Ph\'enom\`enes  Quantiques, 75013  Paris,  France}

\begin{abstract}
Stacking three monolayers of graphene with a twist generally produces two moiré patterns. A moiré of moiré structure then emerges at larger distance where the three layers periodically realign. We devise here an effective low-energy theory to describe the spectrum at
distances larger than the moiré lengthscale. In each valley of the underlying graphene, the theory comprises one Dirac cone at the $\bGM$ point of the moiré Brillouin zone and two weakly gapped points at $\bKM$ and $\bKM'$. The velocities and small gaps exhibit a spatial dependence in the moiré-of-moiré unit cell, entailing a non-abelian connection potential which ensures gauge invariance. The resulting model is numerically solved and a fully connected spectrum is obtained, which is protected by the combination of time-reversal and twofold-rotation symmetries.
\end{abstract}

\maketitle

\section{Introduction}

Following the theoretical prediction and experimental discovery of tantalizing physical properties in twisted bilayer graphene (TBG)~\cite{lopes2007graphene, Suarez2010Flat, bistritzer2011moire, bistritzer2011moirebutterfly, lopes2012continuum, cao2018unconventional, cao2018correlated, carr2019exact, zhang2019landau}, different multilayer stackings of individual graphene sheets or graphene sheets with hexagonal boron nitride layer have been explored where exotic electronic properties may emerge with or without the relative rotations between the layers 
\cite{zhang2011experimental, dean2013hofstadter, burg2019correlated, Khalaf2019Magic, sharpe2019emergent, cao2020tunable, chen2019signatures, hazra2019bounds, tomarken2019electronic, serlin2020intrinsic, chen2020tunable, shen2020correlated, liu2020tunable,wang2020correlated, carr2020ultraheavy, wong2020cascade, zondiner2020cascade, lopez2020electrical, Zhang2021Correlated, zhang2022promotion, chen2021electrically,Xu2021Tunable, shi2021moire, cao2021ab, Lei2021Mirror, park2021flavour, he2021symmetry, xie2021weak, khalaf2021charged, liu2022isospin, eaton2022renormalized}.
With two layers, a geometrical moiré pattern emerges with a periodicity largely exceeding the original unit cell of graphene and kinetic energy is quenched in the vicinity of the magic angle $\sim 1.05^{\circ}$. Experiments have revealed numerous appealing properties such as superconductivity \cite{cao2018unconventional, Yankowitz2019Tuning, lu2019superconductors, stepanov2020untying, saito2020independent, khalaf2021charged}, correlated insulator phase \cite{cao2018correlated, Yankowitz2019Tuning, lu2019superconductors, wang2020correlated, khalaf2021charged}, nematicity \cite{Kozii2019nematic, Chichinadze2020Nematic, lothman2022nematic, Onari2022Mechanism}, integer and fractional Chern insulators \cite{sharpe2019emergent, Ledwith2020Fractional, nuckolls2020strongly, Repellin2020Chern, xie2021fractional, Bergholtz_2021,Sheffer_2021}, spontaneous flavor polarization \cite{Potasz2021Exact, park2021flavour, xie2021weak}, orbital ferromagnetism \cite{he2020giant, Li2020Experimental, Tschirhart2021Imaging,Guerci2021Moire} or strange-metal behavior \cite{polshyn2019large,Wu2019Phonon,Cao2020Strange,jaoui2022quantum}.


\begin{figure}[!ht]
    \centering
    \includegraphics[width=0.9\linewidth]{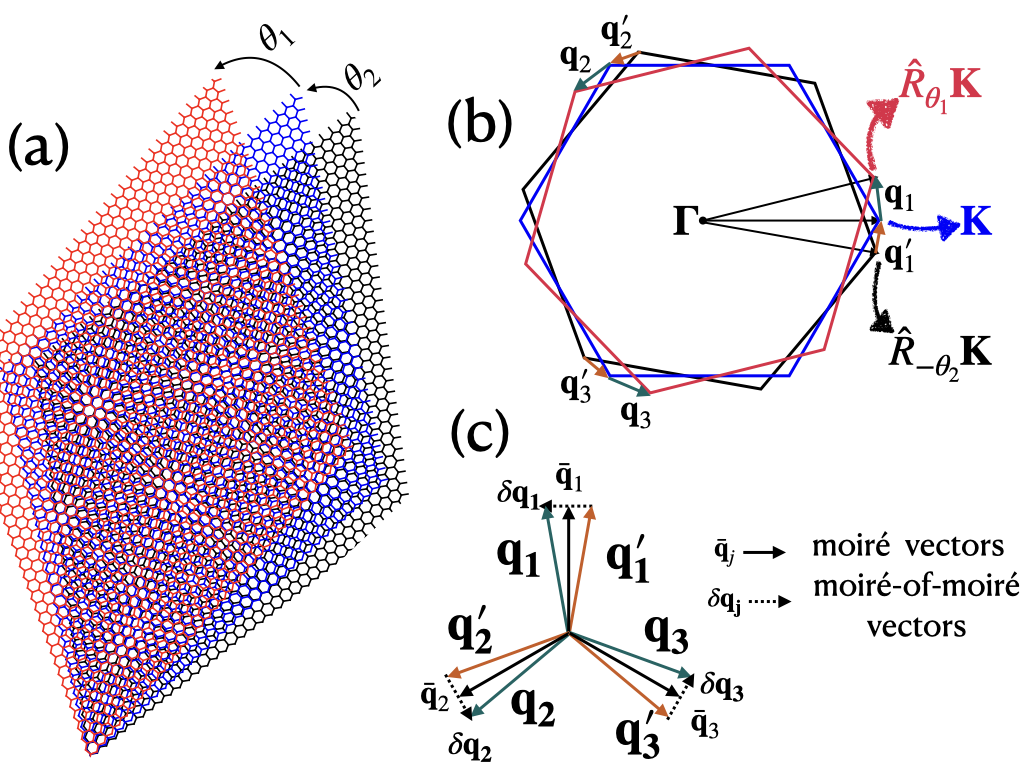}
    \caption{(a) Real-space arrangement of the staircase trilayer graphene monolayers. (b) The Brillouin zones of the 3 graphene sheets. (c) Illustration of the decomposition of native moiré vectors into approximate moiré vectors and super\mr vectors.}
    \label{fig:trilayer}
\end{figure}




Adding a third graphene sheet on top of TBG with a relative rotation generates the twisted trilayer graphene (TTG) configuration. Experiments in TTG have shown evidence of correlated insulating phases and robust superconductivity with unconventional pairing~ \cite{lopez2020electrical, park2021tunable, cao2021pauli, Hao2021Electric, Zhang2021Correlated, kim2022evidence, liu2022isospin}.
In comparison with TBG, TTG exhibits better tunability of the electronic structure in mainly \emph{three} aspects:
(1) the two independent twist angles of TTG allow for an extra degree of freedom of manipulation \cite{zhu2020twisted};
(2) in contrast with TBG where an horizontal shift of one layer can be canceled by an appropriate choice of origin or equivalently, by a unitary transform, such invariance does not always exist in TTG; (3) the band structure of TTG is largely tunable by applying a perpendicular electric field~\cite{chen2019evidence,guerci2021higherorder}.
Experimentally so far, the most successful implementation of TTG~\cite{park2021tunable,cao2021pauli,Hao2021Electric,kim2022evidence} is the symmetric stacking where the top and bottom layers are rotated in the same direction with respect to the middle layer.
Such an arrangement has exactly the same moir\'e periodicity as TBG but depends on the relative horizontal shift of one layer.  Its success lies in a remarkable band structure occurring at a magic angle $\sim 1.48^\circ$ ($\sqrt 2$ times the magic angle of TBG) \cite{Khalaf2019Magic}
where an almost flat band coexists with a large velocity Dirac cone~\cite{Khalaf2019Magic,Li2019Materials,Phong2021Band,zhang2022promotion}.
The correlated phases of symmetric TTG have been explored in a number of theory papers~\cite{PhysRevB.103.195411,PhysRevB.104.115167, christos2022correlated,turkel2022orderly}.

In this paper we focus on yet another configuration: the staircase TTG~\cite{PhysRevLett.123.026402} where the top and the bottom layers are rotated oppositely with respect to the middle layer as illustrated  in Fig.\ref{fig:trilayer}. In this case, the two moiré patterns resulting from pairs of subsequent layers combine to form a supermoiré or moiré of moiré structure~\cite{zhu2020twisted}, even for equal twist angles. The unit cell of supermoiré is parametrically larger than the individual moirés. The corresponding Brillouin zone is in turn much smaller with fewer electron densities than TBG or symmetric TTG. Remarkably, an experimental realization~\cite{Zhang2021Correlated} of staircase TTG has revealed correlated insulating states at electronic densities as small as $10^{10}$ cm$^{-2}$ in agreement with the formation of supermoiré mini-bands which spontaneously choose a spin/valley polarization. The same experiment~\cite{Zhang2021Correlated} also measured zero-resistance states suggestive of superconductivity. The superconducting instability is generally favored by van Hove singularities or by an accumulation of states at certain energies as for instance with flat bands. A strongly peaked density of states at charge neutrality was predicted theoretically~\cite{zhu2020twisted} in the parameter regime of the experiment~\cite{Zhang2021Correlated}. The physical origin behind the accumulation of low-energy states in this model is however unclear.

The aim of our work is to provide an effective low-energy theory for the moiré of moiré band structure. The direct numerical solution to the moiré continuum (or Bistritzer-MacDonald) model extended to TTG can be achieved ~\cite{Amorim2018Electronic,zhu2020twisted} but it requires the diagonalization of large matrices and poses a challenge, for instance with the inclusion of interactions. Our approach is a low-energy limit to the Bistritzer-MacDonald (BM) theory.
Whereas the graphene lattice is coarse-grained in BM to achieve an effective model valid at moiré scales, we further perform a low-energy coarse-graining at the moiré scale and derive an effective, but asymptotically exact, theory for moiré of moiré mini-bands. 

At moiré length scales, staircase TTG has a moiré spectrum~\cite{PhysRevLett.123.026402}, in fact two with the time-reversed valleys ${\bf K}$ and ${\bf K}'$ of the Graphene Brillouin zone. In each valley, we identify three low-energy sectors in the vicinity of the $\bGM$, $\bKM$ and $\bKM'$ points, which makes six sectors in total. Each sector presents a Dirac cone ($\bGM$) or weakly gapped Dirac cone ($\bKM$ and $\bKM'$) where velocities and gaps vary over the supermoiré unit cell. Importantly, this spatial dependence, akin to a motion in a curved space, imposes the emergence of a non-abelian real space connection to ensure the gauge invariance of the model. The impact of the connection on the band structure is crucial: the density of states of the resulting model differs strongly from an average over the different moiré hamiltonians. The connection, in deep analogy with a magnetic field, redistributes weights and tends to regroup energies.

Our approach can be seen as a generalization of the ${\bm k} \cdot {\bm p}$ theory to a smooth change of band structure, a method pioneered by Kohn and Luttinger~\cite{Luttinger1955motion} for semiconductors (see also Ref.~\cite{en1952motion}) and later adapted, under the name of envelope functions, to solve the spectrum of semiconductor heterostructures and superlattices~\cite{bastard1981theoretical,whiteprl1981,bastard1982theoretical,smith1990,Bastard1992}. The specificity of graphene material studied in this work is the presence of Dirac cones as opposed to the quadratic dispersion of semiconductors, and the strong role played by the non-abelian connection amplified by the proximity of the conduction and valence bands.
Our work has also several analogies with conical intersections in the Born-Oppenheimer approximation for molecules~\cite{bohm2003geometric,baer2006beyond}.

In this paper, we derive an effective low-energy model for the super\mr length scale, which consists generically of a space-dependent anisotropric Dirac cone and a non-abelian connection, valid for arbitrary twist angles $\theta_1$ and $\theta_2$ provided that $\theta_1\simeq\theta_2\ll1$. The approach also provides a general framework to derive a low-energy theory problem featuring perturbed periodicities. 
The low-energy spectrum is drastically modified by the super\mr scale which introduces mini-bands close to charge neutrality making the system susceptible to collective instabilities at carriers densities much lower than the \mr ones~\cite{Zhang2021Correlated}. 


We structure the paper as follows. In Sec.~\ref{sec:staircase_continuum_model} we introduce the continuum model for the staircase trilayer graphene. This is followed by the definition in Sec.~\ref{sec:local_H} of the local Hamiltonian which depends parametrically on the super\mr position. The derivation of the effective model is discussed in Sec.~\ref{sec:effective_H}. We discuss the $k\cdot p$ theory in Sec.~\ref{subsec:kpmethod}, the low-energy effective model in Sec.~\ref{subsec:lowenergy_model} and the symmetries of the effective model in Sec.~\ref{subsec:symmetries}. Details on the properties of the anisotropic Dirac cone and of the electronic spectrum are given in Appendix~\ref{app:velocity} and~\ref{app:fullconnection}-\ref{app:ph_symmetry}, respectively. We then present our results for the low-energy electronic spectrum obtained in the case of equal twist angles in Sec.~\ref{sec:bands_equal}. The spectrum for the case $\theta_1/\theta_2=1/2$ is given in Sec.~\ref{sec:bands_ration05}. Appendix~\ref{app:1Dtoy} contains a useful 1D toy model which details the origin of the wavefunction singularity when the effective Dirac velocity changes sign. Finally, we conclude the discussion in Sec.~\ref{sec:conclusions} by briefly summarizing the scope of the work and setting the context for future studies.   

\section{Staircase trilayer graphene}
\label{sec:staircase_continuum_model}

We model three overlaid monolayers of graphene slightly twisted with respect to each other. We consider a staircase configuration (all twist angles are positive when measured from bottom to top) with an angle $\theta_1$ between the middle and top layers and $\theta_2$ between the bottom and middle. 
Twisted trilayer graphene shown in Fig.~\ref{fig:trilayer}(a) develops two moir\'e structures modulated by $\theta_1|\bf K|$ and $\theta_2|\bf K|$ with $\bf K$ the Dirac point of graphene. The mismatch between these two modulations displayed in Fig.~\ref{fig:trilayer}(b) creates an additional super\mr lattice with a weaker modulation. Without loss of generality, assuming $\theta_1/\theta_2\simeq p/q$ with $p$ and $q$ coprime integers, the two \mr wave vectors $\bq_1$ and $\bq'_1$ in Fig.~\ref{fig:trilayer}(c) can be expressed as:
\begin{equation}
\label{moire_of_moire}
    \begin{split}
        &\bq_1=p\bar\bq_1+\delta\bq_1/(p+q),\\
        &\bq'_1=q\bar\bq_1-\delta\bq_1/(p+q)
    \end{split}
\end{equation}
where $\bar\bq_1$ is the wave vector of the \mr pattern while $\delta\bq_1$ is the super\mr modulation which takes into account deviations from perfect commensuration and the lack of collinearity between $\bq_1$ and $\bq'_1$. As long as the two twist angles are small the two wave vectors are well separated $|\delta\bq_1|\ll|\bar\bq_1|$. Eq.~\eqn{moire_of_moire} becomes particularly simple for equal twist angle $\theta_1/\theta_2\simeq1$. In this case the \mr interference pattern develops with a space modulation $\bar\bq_1= \bar{\theta} \, {\bf u}_z \wedge {\bf K}$ with ${\bf u}_z$ axis perpendicular to the TBG plane, $\bar{\theta}=(\theta_1+\theta_2)/2$ and the super\mr wave vector is 
\begin{equation}
\delta {\bq}_1 = \bar{\theta}^2 \,  {\bf u}_z \wedge {\bf u}_z \wedge {\bf K}+\delta \theta \, {\bf u}_z \wedge {\bf K}
\end{equation}
where $\delta\theta=\theta_2-\theta_1$
is the angle deviation. The\mr and super\mr lattices are both triangular. Their principal axis are rotated by $90^\circ$ with respect to each other in the equal-angle case $\delta \theta=0$ and parallel in the opposite limiting case of $\bar{\theta}^2 \ll \delta \theta$. It is worth stressing that, 
as the vectors $\bq_1$ and $\bq'_1$ can never be arranged to be collinear, 
 the \mom pattern develops even when the twist angles are exactly equal $\theta_1 = \theta_2$.

The single-particle band spectrum is described within a continuum model where the Dirac cones in each layer are coupled by the transverse tunneling of electrons. In the K valley, it reads
\begin{equation}
\label{H_trilayer_staircase}
    H_K =\begin{pmatrix}
    \hat{\bk}\cdot\bm \sigma_\theta & \alpha \sum_{j=1}^{3}T_j\,e^{-i\bq_j\cdot\br} & 0\\
    h.c. & \hat{\bk}\cdot\bm \sigma & \alpha \sum_{j=1}^{3}T_j\,e^{-i\bq^\prime_j\cdot\br} \\
    0 & h.c. & \hat{\bk}\cdot\bm \sigma_{-\theta}
    \end{pmatrix},
\end{equation}
whereas the other valley $K'$ is the time-reversal partner $H_{K'} = H_K^*$. We have introduced the differential operator $\hat{\bk} = -i (\partial_x, \partial_y)$, $\bm \sigma = (\sigma_x, \sigma_y)$ and the matrices ($j=0,1,2$)
\begin{equation}
 T_{j+1} = r \sigma_0 + \cos ( 2 \pi j/3) \, \sigma_x  + \sin ( 2 \pi j/3) \, \sigma_y,
\end{equation}
${\bm \sigma}_\theta = e^{i  \sigma_z \theta /2} {\bm \sigma} e^{-i  \sigma_z \theta /2}$, $\alpha=w_{AB}/E_\theta$, with $E_\theta=\hbar v_0 k_D\, \theta$ measuring the dimensionless coupling between the layers. $r$ is the corrugation parameter~\cite{koshino2018}. 
The model is characterized by a set of symmetries. It is invariant under a $2 \pi/3$ rotations around the $z$ axis noted $C_{3 z}$, as well as under $\pi$ rotations around the $x$ axis $C_{2 z}$ when $\theta_1 \simeq \theta_2$. In addition, the symmetry $C_{2 z} \mathcal{T}$ combining time-reversal and $\pi$ rotation around $z$ is also satisfied. In contrast, the particle-hole symmetry, which emerges at the level of the Dirac cones of monolayer graphene, is broken by the Hamiltonian in Eq.~\eqref{H_trilayer_staircase} even when the $\theta$-dependence of the matrices ${\bm \sigma}_\theta$ is neglected. We use the approximation $\sigma_\theta \simeq \sigma$ in the rest of this paper, we remark that this does not qualitatively change our result and simplifies the formulation of our approach. 

\section{Local Hamiltonians}
\label{sec:local_H}

Without loss of generality but for simplicity, in this Section and in the next Sec.~\ref{sec:effective_H} and~\ref{sec:bands_equal} we focus  on the case of almost equal angles $\theta_1 \simeq \theta_2$ to construct our low-energy approach. The case of unequal angles will be treated in Sec.~\ref{sec:bands_ration05}. 

We set the stage for a low-energy approach by identifying a slowly varying potential in the Hamiltonian. 
We decompose the \mr momenta $\bq_j = \bar{\bq}_j + \delta \bq_j/2$, $\bq_j' = \bar{\bq}_j - \delta \bq_j/2$ and rewrite the Hamiltonian as
\begin{equation}\label{H_trilayer_staircase2}
    H_K =\begin{pmatrix}
    \hat{\bk}\cdot\bm \sigma & \alpha V_1 ({\bf r},\phi({\bf r})) & 0\\
    h.c. & \hat{\bk}\cdot\bm \sigma & \alpha V_1 ({\bf r},-\phi({\bf r}))\\
    0 & h.c. & \hat{\bk}\cdot\bm \sigma
    \end{pmatrix}
\end{equation}
with the interlayer hopping potential $V_1 ({\bf r},\phi({\bf r})) = \sum_{j=1}^{3} T_j\,e^{-i\phi_j  ({\bf r})}\, e^{-i\bar{\bq}_j\cdot\br}$. The slow variables are the phases $\phi_j ({\bf r}) = \delta \bq_j \cdot \br/2$ which vary over the \mom superlattice lengthscale. On shorter scales, we can approximate them to be constant $\phi_j ({\bf r}) \simeq \phi_j$. We use in fact the notation $\phi_j ({\bf R})$ to indicate that they are locally constant but  still depend on the position ${\bf R}$ in the \mom superlattice. With this approximation, the Hamiltonian Eq.~\eqref{H_trilayer_staircase2} becomes a set of local Hamiltonians 
\begin{equation}\label{H_trilayer_staircase3}
        H_M ({\bm \phi}) =\begin{pmatrix}
        \hat{\bk}\cdot\bm \sigma & \alpha V_1 ({\bf r},{\bm \phi}) & 0\\
        h.c. & \hat{\bk}\cdot\bm \sigma & \alpha V_1 ({\bf r},-{\bm \phi})\\
        0 & h.c. & \hat{\bk}\cdot\bm \sigma
    \end{pmatrix}
\end{equation}
where the three phases $\bm{\phi} = (\phi_1, \phi_2,\phi_3)$ depend on the position ${\bf R}$ and evolve over the \mom superlattice. In contrast with the exact original Hamiltonian Eq.~\eqref{H_trilayer_staircase}, the local Hamiltonian $H_M ({\bm \phi})$, with fixed phases ${\bm \phi}$, is exactly periodic on the \mr real-space lattice. Similarly to the twisted bilayer case, it is readily solved numerically in Fourier space~\cite{bistritzer2011moire}. Each eigenstate wavefunction expands over an hexagonal lattice in \mr momentum space and the convergence is exponentially fast with the number of lattice points~\cite{bernevig2021}. $H_M ({\bm \phi})$ also possesses a gauge invariance: its spectrum is invariant under a global shift of all phases $\phi_j \to \phi_j + \phi_0$ with arbitrary $\phi_0$. In other words, the first phase $\phi_1$ can always be chosen to be zero. Physically, the phases $(\phi_2,\phi_3)$ can be interpreted as indicators of the local relative position of the middle layer with respect to the top and bottom layers. It is always possible to find a location, then defined as the origin ${\bf r}=0$, where the top and bottom layers align along an $AA$ stacking. We denote two particular cases of interest: (i) $AAA$ local stacking for $(\phi_2,\phi_3)=(0,0)$ and (ii) $ABA$ local stacking for $(\phi_2,\phi_3)=\pm (2 \pi/3,-2 \pi/3)$.

\begin{figure}[!ht]
    \centering
    \includegraphics[width=\linewidth]{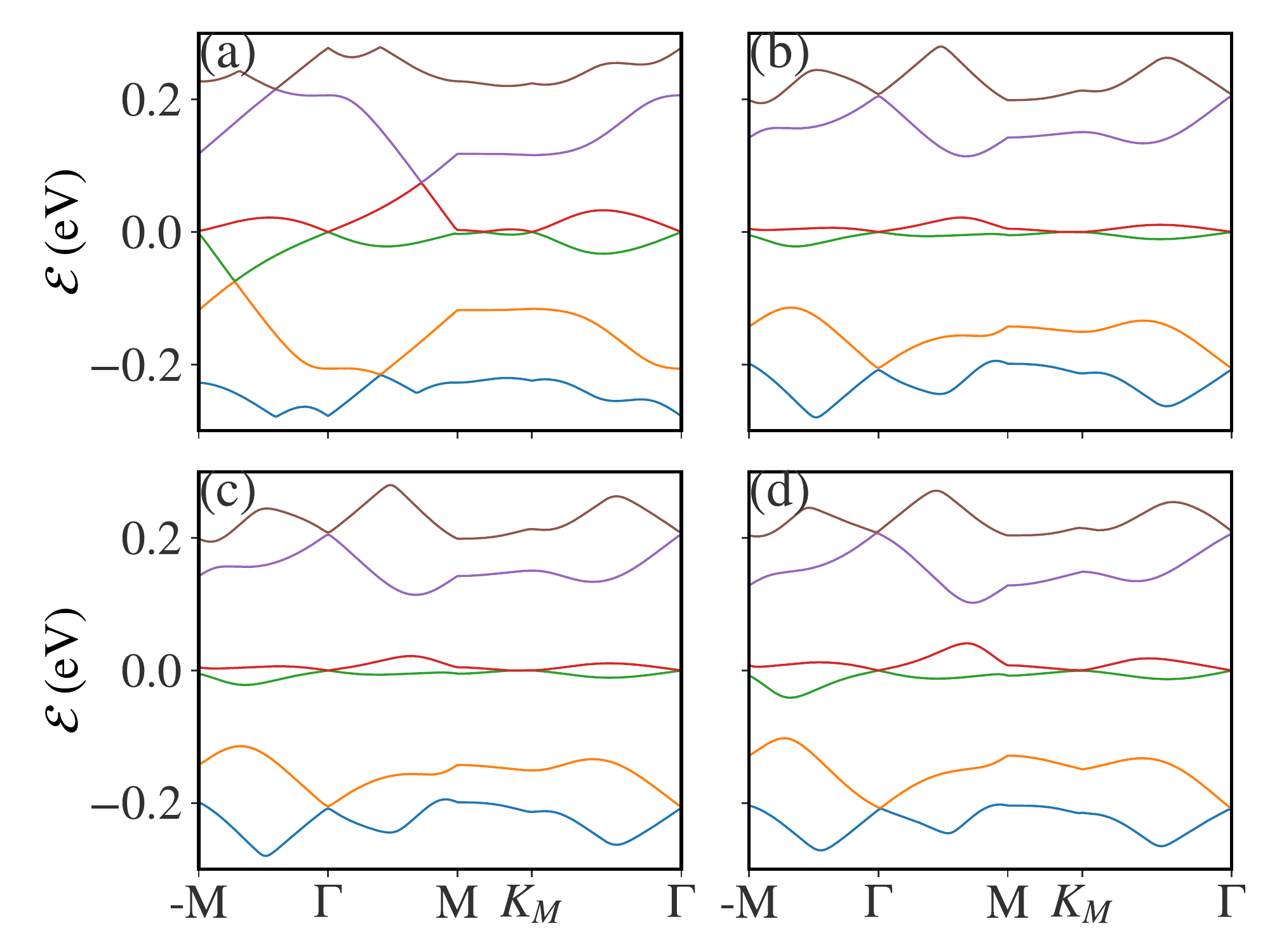}
    \caption{Local spectra for (staircase-wise) twisted trilayer graphene with $\theta_1 \simeq \theta_2 = 2.0^\circ$ calculated for different stacking configurations. The phases $\phi_{2,3}$ map the shifted position of the middle layer before the twisting.
        (a) $(\phi_2, \phi_3) = (0,0)$, AAA stacking (b) $(\phi_2, \phi_3) = (2\pi/3,-2\pi/3)$, ABA stacking (c) $(\phi_2, \phi_3) = (-2\pi/3,2\pi/3) $, ABA stacking (d) $(\phi_2,\phi_3) = (0.4\pi, -0.4\pi)$, generic stacking. }
    \label{fig:local spectra}
\end{figure}

A symmetry analysis reveals that the local Hamiltonian $H_M ({\bm \phi})$ generally breaks all symmetries ($C_{3z}$, $C_{2 x}$ and $C_{2 z} \mathcal{T}$) except at fine-tuned values of $\bm{\phi}$ corresponding to the local $AAA$ and $ABA$ stacking mentioned above. In contrast with the exact Hamiltonian Eq.~\eqref{H_trilayer_staircase}, the local Hamiltonian $H_M ({\bm \phi})$ exhibits a particle-hole symmetry protecting a Dirac cone at $\bGM$, the ${\bf \Gamma}$ point of the \mr Brillouin zone, as proven in Appendix~\ref{app:ph_symmetry}. In the two cases of $AAA$ and $ABA$ stacking, the symmetries $C_{3z}$ and $C_{2 x} C_{2 z} \mathcal{T}$ are preserved protecting one Dirac cone at $\bKM$ and another at $\bKM'$. Aside from these symmetric configurations, a general value of $\bm{\phi}$ weakly gaps the Dirac points at $\bKM$ and $\bKM'$, the larger the angle $\bar{\theta}$ the smaller the gaps. 
In Fig.~\ref{fig:local spectra} the band structures of the local Hamiltonians with different typical phase factors are shown. The case where $\phi_2 = \phi_3 = 0$ (Fig.~\ref{fig:local spectra}(a)) corresponds to the model previously discussed in Ref.~\cite{PhysRevLett.123.026402}.

\section{Effective model}
\label{sec:effective_H}

What we learned from the above analysis in Sec.~\ref{sec:local_H} is that low-energy states, {\it i.e.} states with energies close to charge neutrality, are found in the vicinities of the $\bGM$, the $\bKM$ and $\bKM'$ points of the \mr Brillouin zone. We have moreover obtained a set of local Hamiltonians and, in an incoherent regime, the spectrum would be composed by the additions of all the corresponding eigenvalues. We wish however to describe states that span coherently the entire \mom space. We need to construct states interpolating continuously the solutions of the local Hamiltonians of Sec.~\ref{sec:local_H}. As we demonstrate below, this can be achieved with a quasi-local approach, similar in spirit to the conventional $\bk \cdot \bp$ method but extending it to a spatially varying perturbation. The issue of gauge invariance is crucial in fixing the structure of the resulting low-energy theory.

\subsection{$k \cdot p$ method}
\label{subsec:kpmethod}

Before presenting our theory, let us briefly review the $\bk \cdot \bp$ approach relevant for  the case where the phases $\bm{\phi}$ are spatially homogeneous and we look for the band spectrum of the local Hamiltonian $H_M ({\bm \phi})$ in the vinicity of $\Gamma_M$. There are two zero-energy states at $\Gamma_M$, solutions of the coupled differential equations
\begin{equation}\label{zero-energy-prev}
    H_M ({\bm \phi}) u_{\Gamma,\beta} (\br) = 0  
\end{equation}
where $\beta=1,2$ labels them and the twofold degeneracy is protected by particle-hole symmetry, see Appendix~\ref{app:ph_symmetry}. Assuming a small enough momentum $|\bk| a_M \ll 1$, where $a_M$ denotes the \mr lattice constant, we use a simple {\it ansatz} for the low-energy eigenstates of $H_M ({\bm \phi})$
\begin{equation}
    \psi_{\bk} (\br) = \left[ f_1 u_{\Gamma,1} (\br) + f_2 u_{\Gamma,2} (\br) \right] e^{i \bk \cdot \br}.
\end{equation}
Using that the plane wave $e^{i \bk \cdot \br}$ is almost constant over a moiré unit cell, we project $H_M ({\bm \phi}) \psi_{\bk} (\br) = \varepsilon_{\bk} \psi_{\bk} (\br)$ over the zero-energy states $u_{\Gamma,1/2} (r)$ and integrate over the unit cell to find
\begin{equation}
H_{\rm eff} (\bk)
\begin{pmatrix}
f_1 \\ f_2
\end{pmatrix} = \varepsilon_{\bk} \begin{pmatrix}
f_1 \\ f_2
\end{pmatrix}
\end{equation}
with the low-energy effective Hamiltonian
$H_{\rm eff} (\bk) = {\bm \Gamma} \cdot \bk $ and ${\bm \Gamma}_{\beta,\beta'} = \langle u_{\Gamma,\beta} | {\bm \sigma} | u_{\Gamma,\beta'} \rangle$, or more precisely the integrals over the moiré unit cell
\begin{equation}
{\bm \Gamma}_{\beta,\beta'} = \int_{MUC} d \br \,  u_{\Gamma,\beta}^* (\br)  {\bm \sigma}  u_{\Gamma,\beta'} (\br).
\end{equation}
Interestingly, we have shown here that the vinicity of $\bGM$ is described by a Dirac cone, generally anisotropic, whose properties are entirely determined by the matrix elements between the two zero-energy solutions. In fact, $$H_{\rm eff} (\bk) = {\bm \Gamma} \cdot \bk $$ can be seen as the (leading) first order term of a Taylor expansion. The second order term vanishes to satisfy particle-hole symmetry and the next non-vanishing term is the third order trigonal wrapping contribution that we will not further explore.

There is another important feature related to gauge invariance. The choice of the two zero-energy states $u_{\Gamma,1/2} (r)$ is somewhat arbritrary as any unitary transformation $U$
\begin{equation}
\begin{pmatrix}
 u'_{\Gamma,1} (\br) \\ u'_{\Gamma,2} (\br)
\end{pmatrix}
= U \begin{pmatrix}
 u_{\Gamma,1} (\br) \\ u_{\Gamma,2} (\br)
\end{pmatrix}
\end{equation}
provides another admissible set of orthogonal zero-energy states but a different low-energy Hamiltonian $H'_{\rm eff} (\bk) = {\bm \Gamma}' \cdot \bk $ with ${\bm \Gamma}' = U {\bm \Gamma} U^+$. A simple gauge transform nevertheless retrieves the original effective Hamiltonian $H_{\rm eff} (\bk) = U^+ H'_{\rm eff} (\bk) U$ and the spectrum is clearly invariant on the particular choice of zero-energy states as expected.

\subsection{Low-energy effective model}
\label{subsec:lowenergy_model}

With this construction in mind, we turn to the case of interest where the phases $\bm{\phi}$ vary over the \mom pattern and introduce again the zero-energy states at $\bGM$ given in Eq.~\eqref{zero-energy-prev}, obtained at fixed values of the phases $\phi$,
\begin{equation}\label{zero-energy}
    H_M ({\bm \phi}) u_{\Gamma,\bm{\phi},\beta} (\br) = 0.  
\end{equation}
We have reinstated here the subscript $\bm{\phi}$ to stress the fact that a different pair of zero-energy states is defined for each $\bm{\phi}$. We then use the following {\it ansatz} for the eigenstates of the original continuum Hamiltonian Eq.~\eqref{H_trilayer_staircase2} (not the local one)
\begin{equation}
    \psi (\br) =  f_1 (\br) \, u_{\Gamma,\bm{\phi}(\br),1} (\br) + f_2  (\br) \, u_{\Gamma,\bm{\phi}(\br),2} (\br) 
\end{equation}
where the functions $f_{1,2} (\br)$ are slowly varying on the \mr scale. The functions $u_{\Gamma,\bm{\phi}(\br),1} (\br)$ vary rapidly on the \mr scale through their explicit dependence on $\br$, but also have a slow envelope through $\bm{\phi}(\br)$. Using the separation of length scales, we project $H_K \psi (\br) = \varepsilon \psi (\br)$ onto the zero-energy states and integrate over the \mr unit cell. The result takes the form
\begin{equation}
H_{\rm eff} 
\begin{pmatrix}
f_1 (\bR) \\ f_2 (\bR)
\end{pmatrix} = \varepsilon \begin{pmatrix}
f_1 (\bR) \\ f_2 (\bR)
\end{pmatrix},
\end{equation}
with the effective Hamiltonian ($\hat{\bk} = -i {\bm \nabla}_{\bR}$)
\begin{equation}\label{eff_hamil}
H_{\rm eff} = \frac 1 2 \{ {\bm \Gamma} (\bR) , \hat{\bk} \} + A(\bR)
\end{equation}
defined with the space-dependent anisotropic Dirac matrix
\begin{equation}\label{gamma}
{\bm \Gamma}_{\beta,\beta'} (\bR) = \int_{MUC} d \br \,  u_{\Gamma,\bm{\phi}(\bR),\beta}^* (\br) \,  {\bm \sigma} \,  u_{\Gamma,\bm{\phi}(\bR),\beta'} (\br),
\end{equation}
or in short notation ${\bm \Gamma}_{\beta,\beta'} = \langle u_{\Gamma,\bm{\phi},\beta} | {\bm \sigma} | u_{\Gamma,\bm{\phi},\beta'} \rangle$
and the non-abelian connection
\begin{equation}\label{nonabelian}
A_{\beta,\beta'} (\bR) = \langle u_{\Gamma,\bm{\phi} (\bR),\beta} | \frac{{\bm \sigma} \cdot {\bm  \nabla}_{\bR} -  {\bm \nabla}_{\bR} \cdot {\bm \sigma}}{2}  | u_{\Gamma,\bm{\phi} (\bR),\beta'} \rangle,
\end{equation}
also integrated over the moiré unit cell. The notation $\bR$ in place of $\br$ indicates that the effective Hamiltonian Eq.~\eqref{eff_hamil} has been coarse-grained over the \mr scale around the average position $\bR$.

The Hamiltonian Eq.~\eqref{eff_hamil} describes the vicinity of the $\bGM$ point in the \mr reciprocal space. The same analysis can be repeated around the $\bKM$ and $\bKM'$ points, the only difference being that the zeroth order does not vanish, {\it i.e.} the two states at $\bKM$ ($\bKM'$) have non-zero energy
\begin{equation}\label{solutionKM}
    H_M ({\bm \phi}) u_{K_M,\beta} (\br) = \varepsilon ({\bm \phi})  u_{K_M,\beta} (\br),
\end{equation}
with $\varepsilon ({\bm \phi}) \ne 0$ in general except at specific high-symmetry points. The two eigenstates of  $H_M ({\bm \phi})$ at $\bKM$ with the energies closest to zero (in practice, they have opposite energy) form a two-dimensional space. Taking an arbitrary basis of this subspace (it does not have to be the basis that diagonalizes $H_M ({\bm \phi})$), we find the effective Hamiltonian
\begin{equation}\label{eff_hamil2}
  H_{\rm eff}^{K_M} =  \mathcal{E} (\bR) + \frac 1 2 \{ {\bm \Gamma} (\bR) , \hat{\bk} \} + A(\bR)
\end{equation}
with the same operators ${\bm \Gamma} (\bR)$ and $A(\bR)$  as above (expect that they are computed within the basis chosen at $\bKM$) together with
\begin{equation}
\mathcal{E}_{\beta,\beta'} (\bR) = \langle u_{K_M,\bR,\beta} | H_{\rm M}  (\bR)  | u_{K_M,\bR,\beta'} \rangle.
\end{equation}
 The same effective model can be written at $\bKM'$.

The effective Hamiltonians Eqs.~\eqref{eff_hamil} and Eq.~\eqref{eff_hamil2} are the central results of our low-energy approach. We emphasize that they cannot be solved locally, that is for a given position $\bR$, as the operator $\hat{\bk}$ does not commute with $\bR$. They are nevertheless readily solved in momentum space, using the Hamiltonian \mom periodicity as discussed in appendix~\ref{app:manual}, and thus describe wavefunctions spanning the entire \mom space.

\subsection{Symmetries}
\label{subsec:symmetries}

We have derived in Eq.~\eqref{eff_hamil} an effective low-energy Hamiltonian for the vicinity of the $\bGM$ point. The model restores all spatial symmetries: $C_{3z}$, $C_{2 x}$ and $C_{2 z} \mathcal{T}$ but breaks particle-hole symmetry. A general proof of the symmetries in the effective model is shown in the Appendix~\ref{app:symmetry}. The  $C_{2 z}\mathcal{T}$ symmetry protects the Dirac point and imposes a fully connected spectrum, {\it i.e.} all bands touch via Dirac points, see Appendix~\ref{app:fullconnection}.

The arbitrary choice for the basis of eigenstates at the $\bGM$ point entails a non-abelian gauge invariance. A different choice of basis,
\begin{equation}\label{eq:local gauge change}
\begin{pmatrix}
 u'_{\Gamma,\bm{\phi}(\bR),1} (\br) \\ u'_{\Gamma,\bm{\phi}(\bR),2} (\br)
\end{pmatrix}
= U (\bR)  \begin{pmatrix}
 u_{\Gamma,\bm{\phi}(\bR),1} (\br) \\ u_{\Gamma,\bm{\phi}(\bR),2} (\br)
\end{pmatrix},
\end{equation}
with the unitary (space-dependent) operator $U (\bR)$, changes the Hamiltonian into $H_{\rm eff}' = U H_{\rm eff} (\bk) U^+$. The wavefunctions are then modified but the spectrum is invariant as the two Hamiltonians are directly related by a unitary transform. The presence of the non-abelian connection of Eq.~\eqref{nonabelian} is crucial in maintaining gauge invariance.

The local Hamiltonian Eq.~\eqref{H_trilayer_staircase3} also possesses translation invariance on the \mom lattice. We introduce the \mom basis vectors ${\bf a}_1^{MM}$, ${\bf a}_2^{MM}$. The corresponding reciprocal basis is formed by${\bf b}_1^{MM} = \delta \bq_2 - \delta \bq_1$, ${\bf b}_2^{MM} = \delta \bq_3 - \delta \bq_1$. The periodicity takes the form
\begin{equation}\label{eq:period local moire}
H_M (\bR + {\bf a}_{1/2}^{MM}) = e^{i \hat{\bk} \cdot {\bf a}_{1/2}^{M}/2} V H_M (\bR) V^\dagger e^{-i \hat{\bk} \cdot {\bf a}_{1/2}^{M}/2}
\end{equation}
with the shorthand notation $H_M (\bR) \equiv H_M ({\bm \phi}(\bR))$ and the diagonal sewing matrix 
$V = \text{diag}(1, 1, e^{2i\pi/3})$.
In Eq.~\eqref{eq:period local moire} ${\bf a}_1^{M}$ and ${\bf a}_2^{M}$ are the generators of the moiré lattice with the reciprocal vectors ${\bf b}_1^{M} = \bar{\bq}_2 - \bar{\bq}_1$ and ${\bf b}_2^{M} = \bar{\bq}_3 - \bar{\bq}_1$.
Since the two Hamiltonians are related by a unitary transformation, their spectra coincide, showing that the local Hamiltonian $H_M (\bR)$ has a periodic spectrum on the triangular lattice generated by ${\bf a}_1^{MM}$ and ${\bf a}_2^{MM}$.
Without loss of generality, we use a specific gauge where the operator $\sigma_z$ (acting the $A/B$ sublattice space) is diagonalized by the basis of eigenstates at $\bGM$. In addition, we require the two states to have a real value for the component on the middle layer when averaged over the \mr lattice. With this choice of gauge, the operators ${\bm \Gamma}(\bR)$ and $A (\bR)$ in Eqs.~\eqref{gamma} and~\eqref{nonabelian} can be shown to be periodic over the \mom lattice which enforces the spectrum of the effective Hamiltonian $H_{\rm eff}$ in Eq.~\eqref{eff_hamil} to be periodic as well.

\section{Band structures for \texorpdfstring{$\theta_1=\theta_2$}{t1=t2} }
\label{sec:bands_equal}

Equipped with our effective Hamiltonians at the $\bGM$, $\bKM$ and $\bKM'$ (corresponding to the valley ${\bf K}$, there are also three time-reversed Hamiltonians at ${\bf K}'$), we can compute the low-energy spectrum. 
Eqs.~\eqref{eff_hamil} and~\eqref{eff_hamil2} are solved by expanding in Fourier components in the \mom reciprocal space where the convergence is exponentially fast with the number of kept momenta.

\begin{figure}[!ht]
    \centering
    \includegraphics[width=0.9\linewidth]{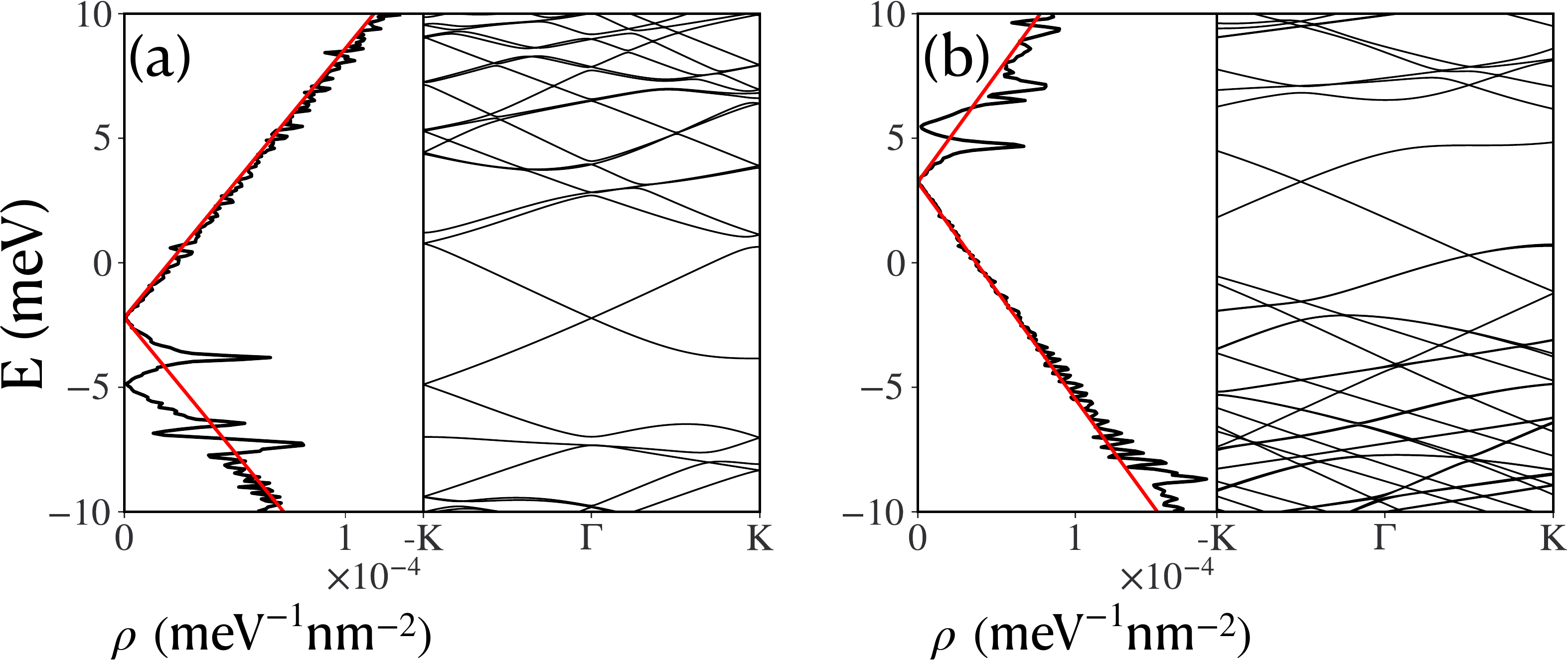}
    \caption{Low-energy band structure and density of states computed using the effective model at (a) $\bGM$ and (b) $\bKM$ points for equal twist angles of $ \theta = 2.2^\circ$. Black lines show the density of states of the effective model smoothed by Gaussian smearing. Red lines show the average density of states of the local (moir\'e) Hamiltonians defined by Eq.~\eqref{eq:mean rho}. The position of the Dirac point is shifted by hand for comparison with the effective model.}
    \label{fig:bands and dos}
\end{figure}

\begin{figure}[!ht]
    \centering
    \includegraphics[width=0.9\linewidth]{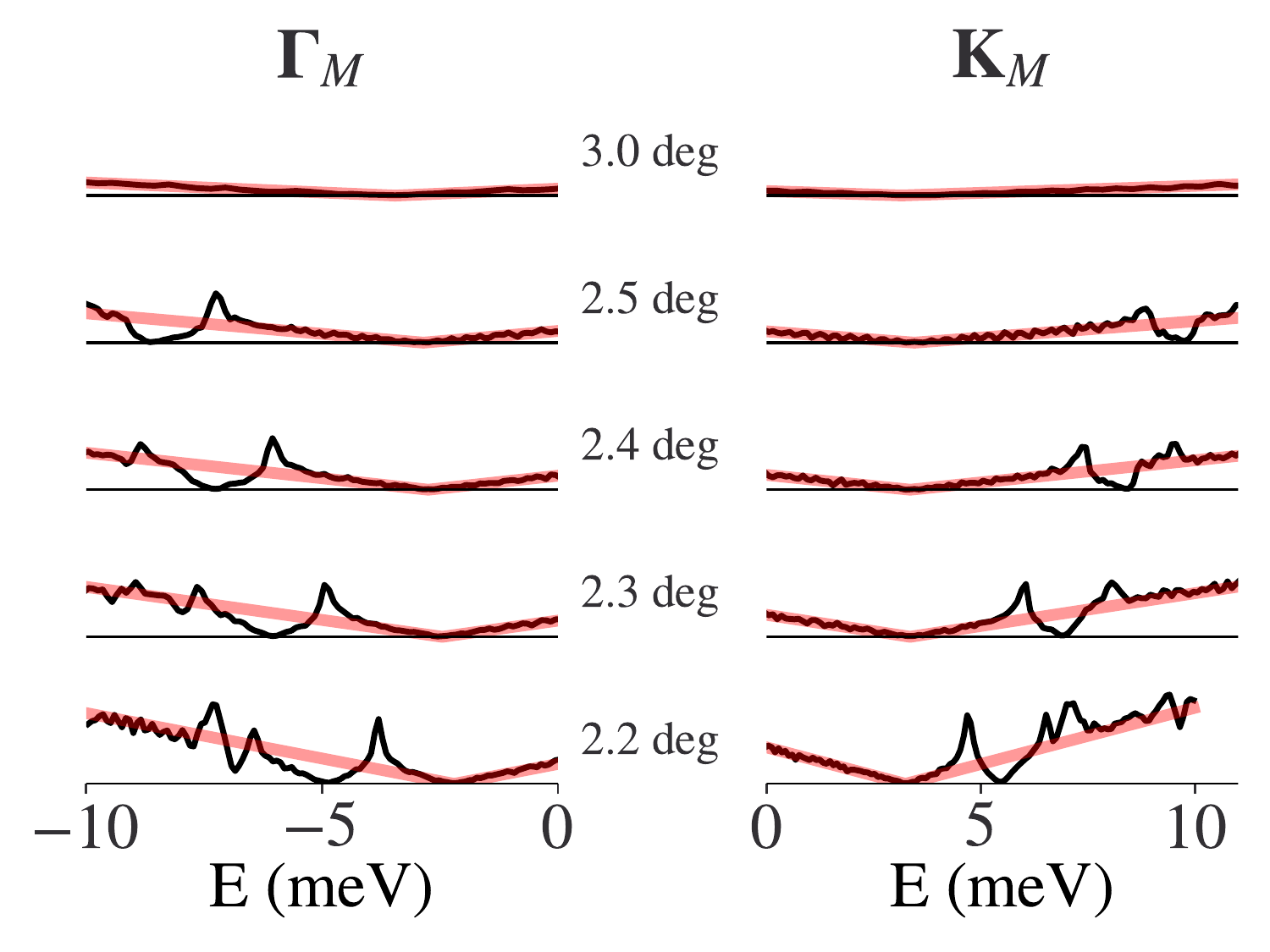}
    \caption{Evolution of the low-energy density of states as a function of the twist angle $\theta_1 =\theta_2$. The red transparent line indicates the density of states averaged over the local Hamiltonian, Eq.~\eqref{eq:mean rho}, with a Dirac point shifting as in Fig.~\ref{fig:bands and dos}.}
    \label{fig:dos eq angles}
\end{figure}

In Fig.~\ref{fig:bands and dos} we show the band structures and density of states computed for the twist angles $\theta = 3.0^\circ$ and $\theta = 2.2^\circ$. The band structures are calculated along the path ${\bf K}-{\bf \Gamma}-{\bf K'}$. 
In Fig.~\ref{fig:dos eq angles} we show the evolution of the low-energy density of states as the symmetric twist angles decrease.

Our first observation for the spectra of Fig.~\ref{fig:bands and dos} is that the energy bands are all fully connected.
This is a consequence of the $C_{2 z} \mathcal{T}$ symmetry~\cite{PhysRevLett.123.026402,ahn2019failure} and the fact that each of the graphene monolayers hosts originally one Dirac cone per valley.
A detailed proof of this property is given in appendix~\ref{app:fullconnection}. We also note that, despite the fact that the dispersion in the values of the Dirac cone velocities at $\bGM$ is weak as $\bR$ varies over the moir\'e of moir\'e lattice, the resulted spectrum exhibits pronounced features with peaks close to energies of vanishing density of states. This behavior is particularly visible in the left panel of Fig.~\ref{fig:bands and dos} computed at $\theta=2.2^\circ$ where we show in black the density of states of the effective model defined in Eq.~\eqn{eff_hamil2}. For comparison, we consider the case where the densities of states of the different local Hamiltonians introduced in Sec.~\ref{sec:local_H} would be simply additive. The resulting average density of states is shown in red in Figs.~\ref{fig:bands and dos} and~\ref{fig:dos eq angles} and it is defined precisely as
\begin{equation}\label{eq:mean rho}
    \bar\rho(\epsilon)=\int\frac{d^2\bR}{\Omega}\rho(\epsilon,\bR),
\end{equation}
where $\Omega$ is the area of the \mom unit cell and $\rho(\epsilon,\bR)$ is the density of states of the local Hamiltonian $H_M( \phi(\bR))$ in Eq.~\eqref{H_trilayer_staircase3}.

Figs.~\ref{fig:bands and dos} and~\ref{fig:dos eq angles} strikingly demonstrate that the exact density of states is far from being the addition of local densities of state. As wavefunctions spread over the full \mom unit lattice and connect the different local Hamiltonian, they redistribute energies. One can also view the formation of  peaks in the density of states as the result of the non-abelian connection $A(\bR)$ in Eq.~\eqref{eff_hamil2}. To some extend, $A(\bR)$ acts as an effective (inhomogenous) vector potential with the tendency to flatten bands and redistribute spectral weights to concentrate it near certain energy levels. 
Although the $C_{2 z} \mathcal{T}$ symmetry excludes the formation of truly separated minibands, we observe almost minibands separated by locations of vanishing density of states where isolated Dirac points form. In the presence of strong Coulomb interactions, it is expected that these minibands are able to form symmetry broken states through collective (Stoner) instabilities~\cite{wong2020cascade,zondiner2020cascade}. We stress that the formation of the pseudo-minibands cannot be captured by averaging over the local \mr models.

\begin{figure}[!ht]
    \centering
    \includegraphics[width=0.9\linewidth]{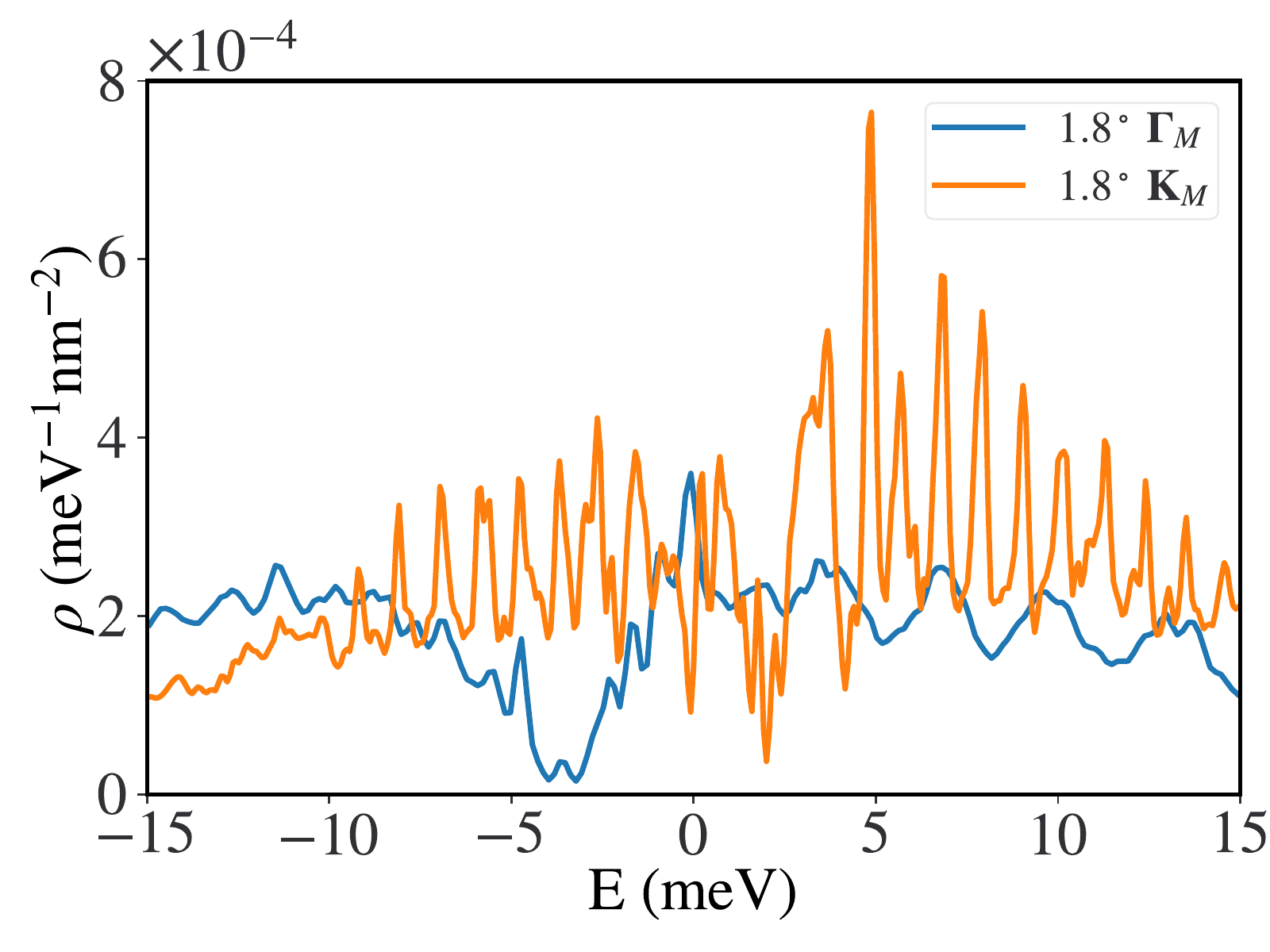}
    \caption{The density of state profiles and band structures for equal twists of $1.8^\circ$ of the effective model calculated at the $\bGM$ and $\bKM$ points.}
    \label{fig:eq dos 1.8 non-conv}
\end{figure}

With the angle dropping below $2^\circ$, the density of states profiles shown in Fig.~\ref{fig:eq dos 1.8 non-conv} are characterized by a strong enhancement of the low-energy density of states caused by the accumulation of many single-particle energy levels. In this regime, the low-energy spectrum does however not converge with the number of kept momenta. It can be 
understood by inspecting the spatial distribution of the Dirac cone velocities  of the local Hamiltonians $H_M(\phi)$ across the \mom unit cell.
The dependence with the high-energy cutoff coincides with a vanishing velocity component at one point in the unit cell, indicating the need for a short-distance regularization, see also Appendix~\ref{app:1Dtoy}. Notably, such a dependence on the high-energy cutoff was also observed in Ref.~\cite{zhu2020twisted} where it was argued that the quantitative value of the density of state was cutoff-dependent but not its main features such as peak positions and peak splitting. 

In our model, the main feature below $2^\circ$ is the absence of a ``V''-shape, characteristic of a Dirac cone at charge neutrality - compare Fig.~\ref{fig:eq dos 1.8 non-conv} with Figs.~\ref{fig:bands and dos} and~\ref{fig:dos eq angles}. The ``V''-shape is replaced by a broad energy distribution corresponding to a state concentration close to charge neutrality. We expect this feature to be insensitive to the details of short-distance regularization.


\begin{figure}[!ht]
    \centering
    \includegraphics[width=\linewidth]{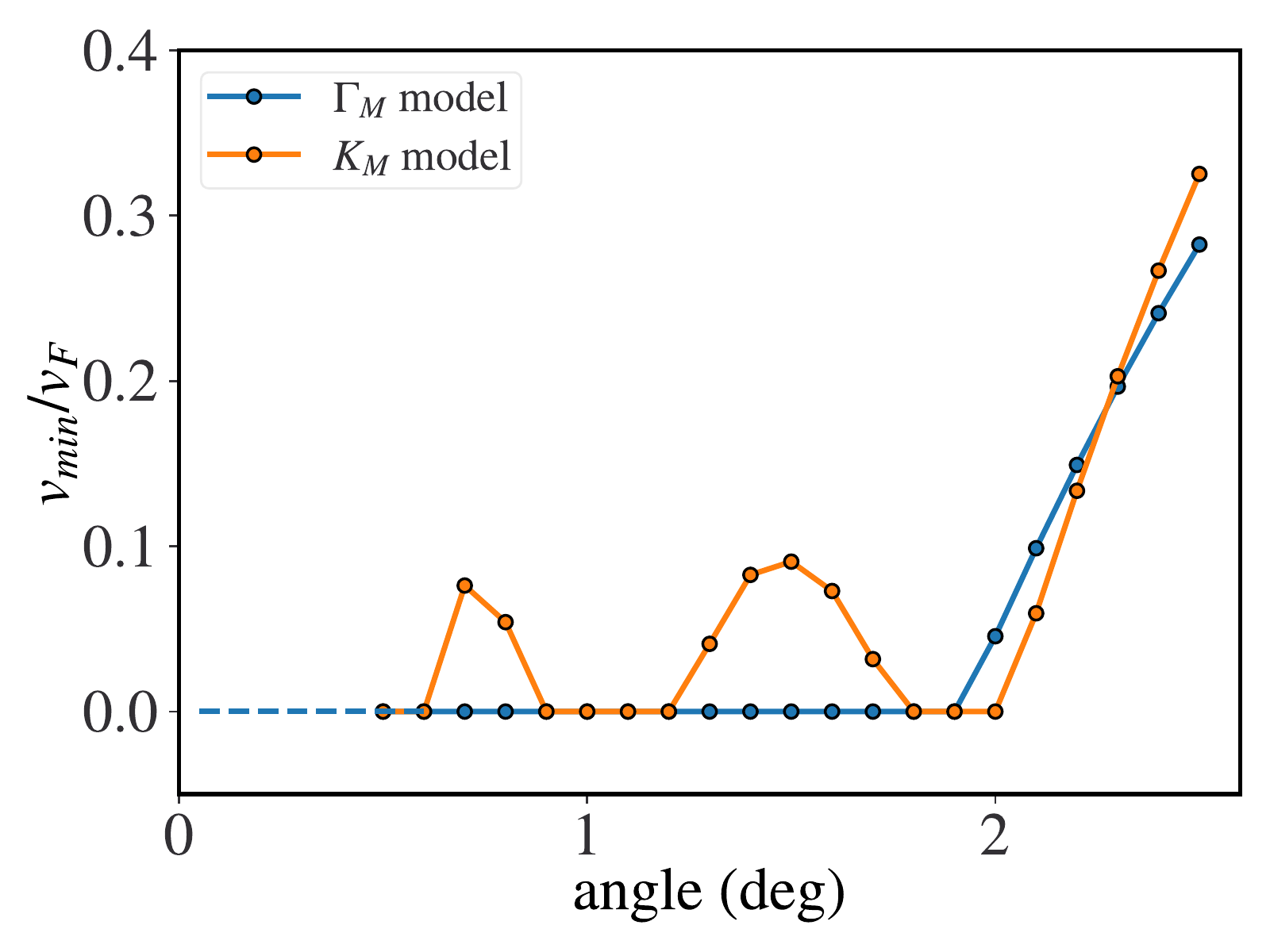}
    \caption{Minimal velocities of the Dirac cones in the local Hamiltonians (see Appendix~\ref{app:velocity}), at $\bKM$ and $\bGM$, as function of the twist angle $\theta_1 = \theta_2$.}
    \label{fig:vmin vs angle}
\end{figure}

In Fig.~\ref{fig:vmin vs angle} we sketch the minimum value of the components of the velocity of the anisotropic Dirac cone found within the \mom unit cell at different symmetrical twist angles. Below $2^\circ$, the $\bGM$ effective model always possesses a vanishing velocity component.
In contrast, the minimal velocity of the $\bKM$ model exhibits revivals at certain ranges of the twist angle. The representation of the two components of an anisotropic Dirac cone is explained in Appendix~\ref{app:velocity}.

\section{Generalization for \texorpdfstring{$\theta_1/\theta_2 \simeq 1/2$}{t1t2}}
\label{sec:bands_ration05}

The central result of this work is the low-energy effective model Eq.~\eqref{eff_hamil2} (and Eq.~\eqref{eff_hamil}) derived in Sec.~\ref{sec:effective_H} in the specific case $\theta_1 \simeq \theta_2$, with band structure calculations in Sec.~\ref{sec:bands_equal}. Eq.~\eqref{eff_hamil2} is in fact much more general and applies similarly to the case of a rational ratio, $\theta_1/\theta_2 \simeq p/q$, with little generalization. The procedure follows closely the one in Sec.~\ref{sec:effective_H}: after identifying a low-energy region in the moiré spectrum of the local Hamiltonians with $n$ bands, for instance $n=2$ at ${\bf \Gamma}_M$ for $\theta_1/\theta_2 \simeq 1$, the Hamiltonian is projected onto $n$ states yielding  Eq.~\eqref{eff_hamil2} with $n\times n$ matrices ${\bm \Gamma} (\bR) $ and $A (\bR)$.

\begin{figure}[!ht]
    \centering
    \includegraphics[width=0.9\linewidth]{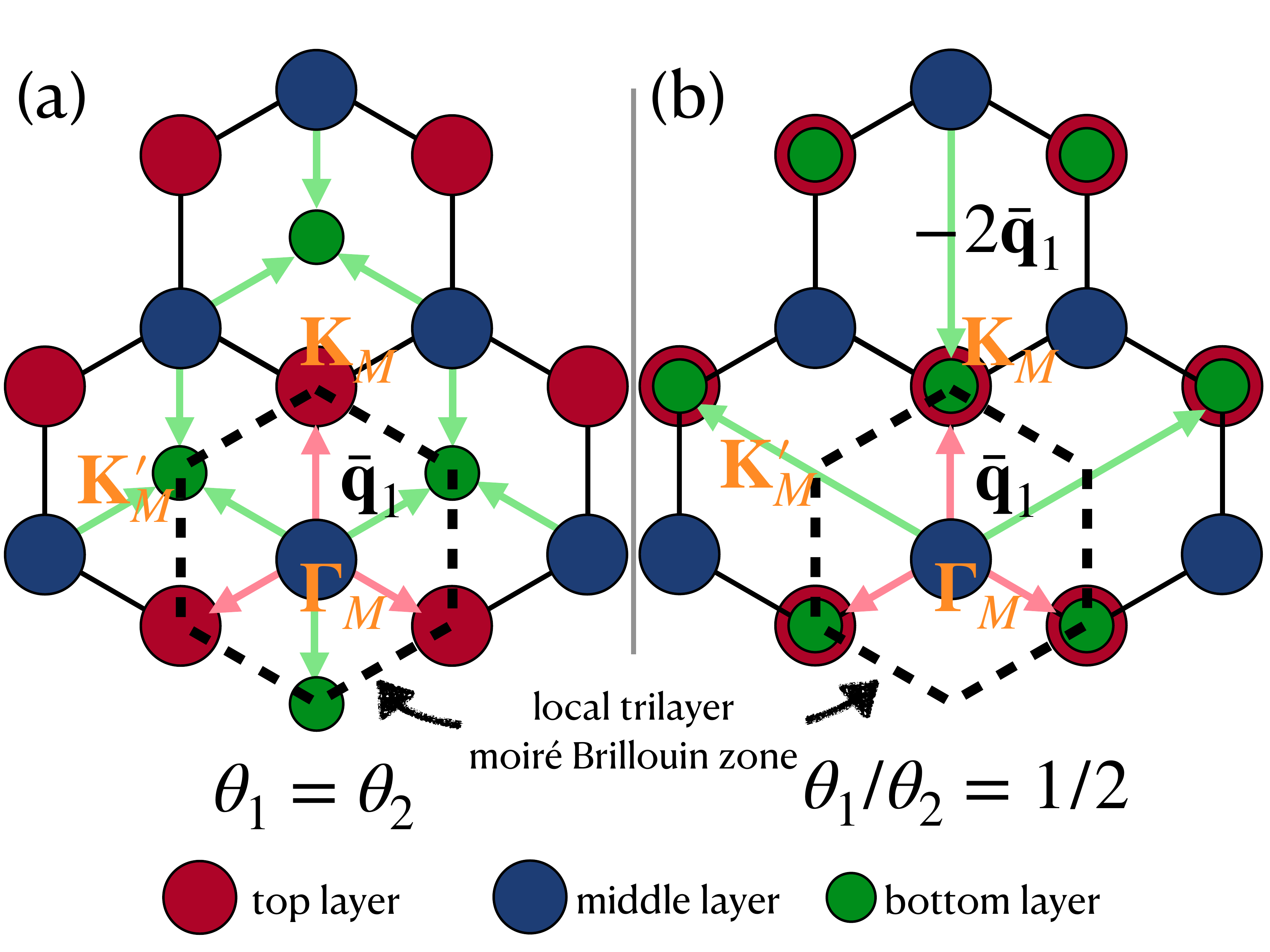}
    \caption{Comparison of the \mr reciprocal lattices for the local Hamiltonians.  
    (a) $\theta_1 = \theta_2$, the three original Dirac cones in the valley $K$ are located at $\bGM$, $\bKM$ and $\bKM'$ in the \mr reciprocal lattice.
    (b) $\theta_1 / \theta_2 = 1/2$, the Dirac cone (valley $K$) of the middle layer is originally located at $\bGM$ while the Dirac cones from the top and bottom layer both overlap at $\bKM$.}
    \label{fig:uneuqal coupling}
\end{figure}

\begin{figure}[!ht]
    \centering
    \includegraphics[width=\linewidth]{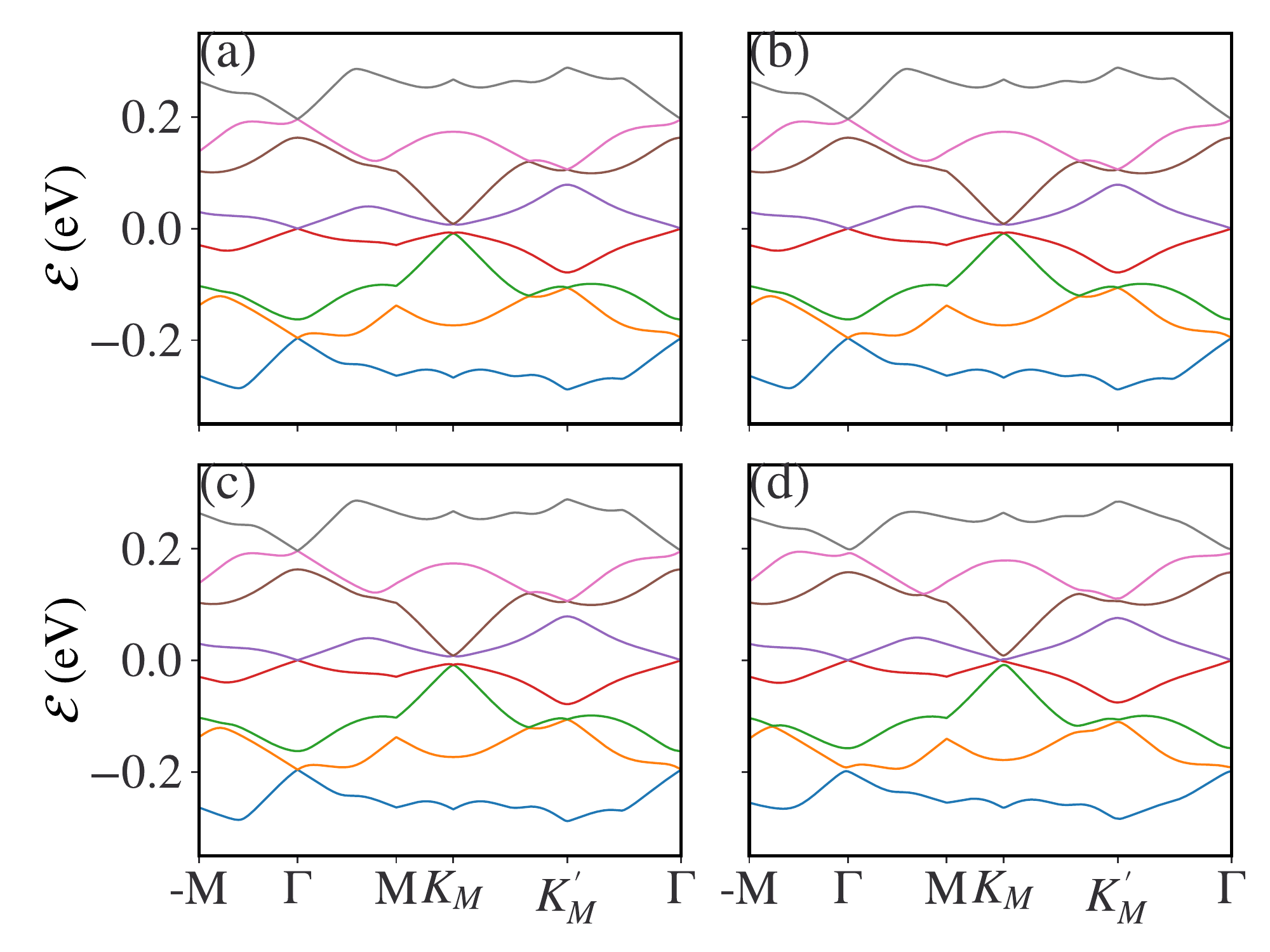}
    \caption{Local \mr band structures computed with $(\theta_1, \theta_2) = (1.8^\circ, 3.6^\circ)$ for different stacking configurations: 
    (a) $(\phi_2, \phi_3) = (0,0)$, AAA stacking 
    (b) $(\phi_2, \phi_3) = (2\pi/3,-2\pi/3)$, ABA stacking
    (c) $(\phi_2, \phi_3) = (-2\pi/3,2\pi/3)$, ABA stacking
    (d) $(\phi_2, \phi_3) = (0.4\pi,-0.4\pi)$, generic stacking.}
    \label{fig:unequal moire band}
\end{figure}

The case $\theta_1/\theta_2 \simeq 1/2$ is the situation explored experimentally in Ref.~\cite{Zhang2021Correlated} and also in Ref.~\cite{zhu2020twisted}. We discuss this regime below as another representative case.
The vectors ${\bq}_1 = \hat R(\theta_1) {\bf K} - {\bf K}$ and ${\bq}'_1 = {\bf K} - \hat R(-\theta_2) {\bf K}$ have different lengths and the decomposition of Eq.~\eqref{moire_of_moire} becomes  $\bq_1 = \bar{\bq}_1 + \delta \bq_1/3$, $\bq'_1 = 2\bar{\bq}_1 - \delta \bq_1/3$.
Starting with the continuum model Eq.~\eqref{H_trilayer_staircase}, the local Hamiltonians at different positions $\bR$ in the supermoiré unit cell take the form
\begin{equation}\label{H_trilayer_staircase2_unequal}
    H_M (\bR)=\begin{pmatrix}
    \hat{\bk}\cdot\bm \sigma & \alpha V_1 ({\bf r},\phi({\bf R})) & 0\\
    h.c. & \hat{\bk}\cdot\bm \sigma & \alpha V_2 ({\bf r},-\phi({\bf R}))\\
    0 & h.c. & \hat{\bk}\cdot\bm \sigma
    \end{pmatrix}.
\end{equation}
$V_1(\bm r, \phi(\bm R))$ has already been defined in Sec.~\ref{sec:local_H} but here $\phi_j(\bR) = \delta \bq_j \cdot \bR / 3$, and $V_2(\br, \phi(\bR)) = \sum_{j=1}^3 T_j e^{-i\phi_j(\bR)} e^{-2i\bar{\bq}_j\cdot \br}$.
The local Hamiltonian Eq.~\eqref{H_trilayer_staircase2_unequal} is diagonalized by expanding the wavefunction over the moiré reciprocal momenta. The corresponding band spectrum evolves periodically in $\bR$ over the supermoiré triangular lattice as a consequence of the identities
\begin{equation}
    H_M (\bR + {\bf a}_{1/2}^{MM}) = e^{i \hat{\bk} \cdot {\bf a}_{1/2}^{M}/3} V H_K (\bR) V^\dagger e^{-i \hat{\bk} \cdot {\bf a}_{1/2}^{M}/3}
\end{equation}
with the \mom basis vectors ${\bf a}_1^{MM}$, ${\bf a}_2^{MM}$. Examples of band structures are shown in Fig.~\ref{fig:unequal moire band} for different stacking configurations. Although the symmetries 
$C_{3 z}$ and $C_{2 z} \mathcal{T}$ apply for the continuum model Eq.~\eqref{H_trilayer_staircase} (but not $C_{2 x}$
when $\theta_1/\theta_2 \simeq 1/2$), these symmetries are broken at the level of the local Hamiltonian, except at fine-tuned $\phi$.
A pair of low-energy states at ${\bf \Gamma}_M$, four at ${\bf K}_M$ and none at ${\bf K}'_M$ can be seen in Fig.~\ref{fig:unequal moire band}, in stark contrast with the case  $\theta_1 = \theta_2$. This can be traced back to the positions of the original single-layer Dirac cones in absence of tunneling, as represented in Fig.~\ref{fig:uneuqal coupling} for $\theta_1 = \theta_2$ and $\theta_1/\theta_2=1/2$. Whereas the three Dirac cones divide over the ${\bf \Gamma}_M$, ${\bf K}_M$ and ${\bf K}'_M$ positions in the former case, there is one Dirac cone at ${\bf \Gamma}_M$ and two at ${\bf K}_M$ in the latter.

Using the moiré solutions of the local Hamiltonians Eq.~\eqref{H_trilayer_staircase2_unequal}, the derivation of the low-energy theory follows the same line as in Sec.~\ref{sec:effective_H}. At $\bGM$, we project onto the two states close to zero energy and recover exactly Eq.~\eqref{eff_hamil2}, except that the energy term $\mathcal{\bm E} (\bR) $ is non-zero here since there is no particle-hole symmetry protecting a Dirac point, see Appendix~\ref{app:ph_symmetry}. The situation is quite different at $\bKM$ where the projection requires four states. We keep the solutions Eq.~\eqref{solutionKM} with $\beta$ running over four values. Smooth wavefunctions with respect to $\bR$ are obtained numerically by the projection technique discussed in Ref.~\cite{Marzari2012Maximally}. We recover again Eq.~\eqref{eff_hamil2} but Eqs.~\eqref{gamma} and~\eqref{nonabelian} are now $4 \times 4$ matrices. We emphasize that the symmetries $C_{3 z}$ and $C_{2 z} \mathcal{T}$ are again restored at the level of the effective theory.

\begin{figure}[!ht]
    \centering
    \includegraphics[width=0.9\linewidth]{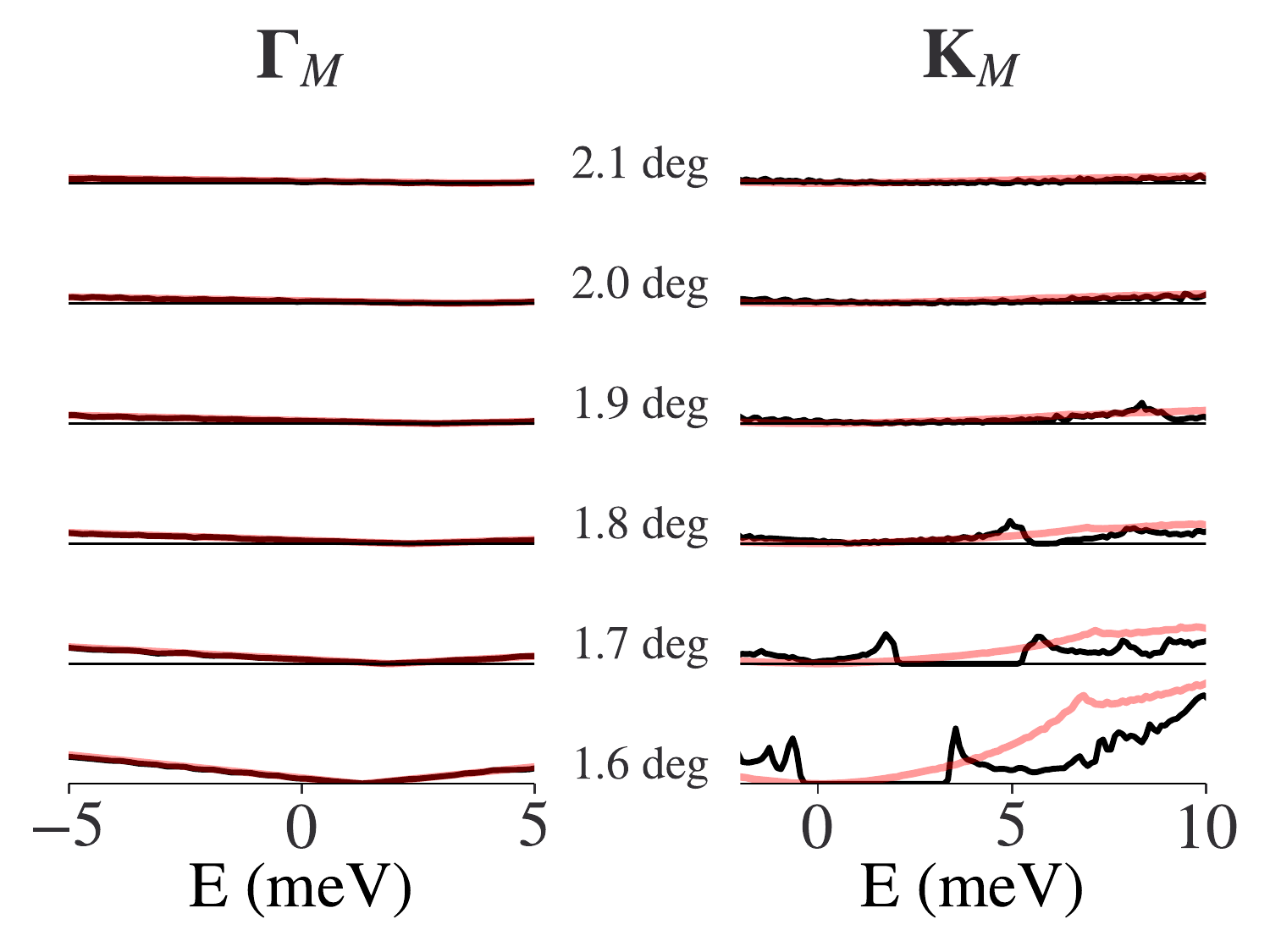}
    \caption{Evolution of the density of states as a function of $\theta_1 = \frac{1}{2}\theta_2$ for the $\bGM$ (left panel) and the $\bKM$ (right panel) effective models. The red lines show the average density of states
    obtained from Eq.~\eqref{eq:mean rho} for comparison - with a proper shift in energy for clarity.
    \label{fig:dos uneq angles}}
\end{figure}

\begin{figure}[!ht]
    \centering
    \includegraphics[width=\linewidth]{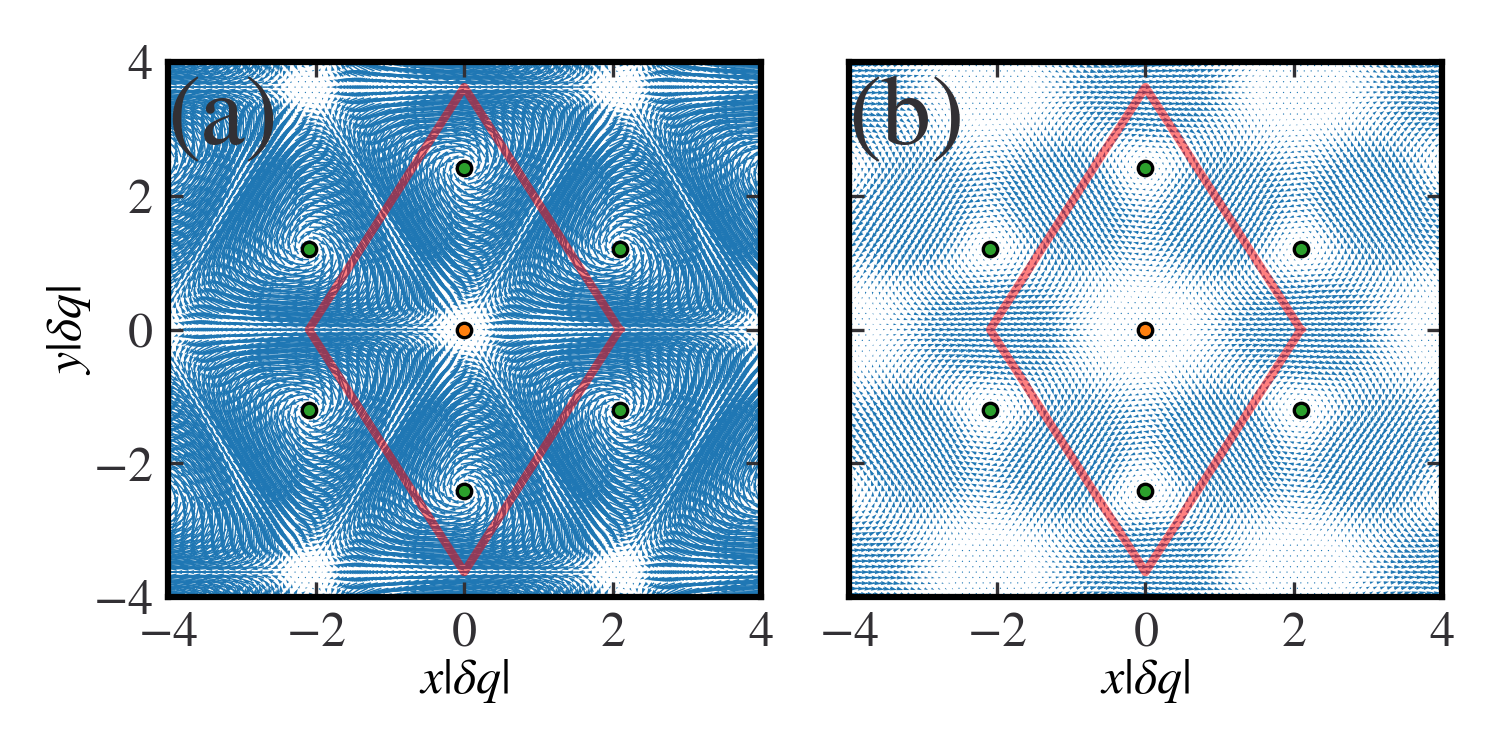}
    \caption{Effective pseudovector potential components $a_x$ 
 and $a_y$ shown as vectors fields in the \mom unit cell delimited by the red diamond. We use the decomposition $A(\bR) = {\bm a} (\bR) \cdot {\bm \tau} + a_0 (\bR) \tau_0$
    for the non-abelian gauge potential. The left and right panels are calculated for (a) $\theta_1 = \theta_2 = 2.1^\circ$ ($\delta q = 0.021~ \text{nm}^{-1}$) and (b) $\theta_1 = \theta_2 /2 = 2.1^\circ$ ($\delta q = 0.069~\text{nm}^{-1}$, vector size increased by a factor $4$ for visibility), with far more structure for equal twist angles. The high-symmetry \mom coordinates where $C_{3z}$ symmetry is restored for the local model are marked by orange and green dots, where the local model has AAA and ABA stacking, respectively. The pseudo-vector potential winds around these high symmetry points.}
    \label{fig:pseudomag field}
\end{figure}

In Fig.~\ref{fig:dos uneq angles} we show the low-energy density of states computed from the effective theory for different values of $\theta_1$ at ${\bf \Gamma}_M$ and ${\bf K}_M$. At ${\bf \Gamma}_M$, the mini-band features observed in Sec.~\ref{sec:bands_equal} are absent and the dispersion essentially coincides with an averaged Dirac cone. 
At low angles, the velocity is suppressed, increasing the slope of the density of states. 
The absence of the mini-bands can be understood by looking at the non-abelian connection $A (\bm R)$ whose curl acts as a pseudo-magnetic field for the electrons.
Fig.~\ref{fig:pseudomag field}(a) and Fig.~\ref{fig:pseudomag field}(b) shows the in-plane components $a_i(\bR)=\Tr[\tau^i A(\bm R)]/2$ with $i=x,y$ of the non-abelian connection for the two different configurations $\theta_1=\theta_2$ and $\theta_1=\theta_2/2$, respectively. We readily realize that the amplitude of the magnetic field is much larger for the equal twist-angle configuration which gives rise to pronounced mini-band features shown in Fig.~\ref{fig:dos eq angles}. On the other hand, for $\theta_1/\theta_2=1/2$ the gauge field is small and the density of states black line in Fig.~\ref{fig:dos uneq angles} is well described by the adiabatic one red line in Fig.~\ref{fig:dos uneq angles}. 

At $\bKM$, the calculated density of states exhibits some interesting features as the twist angle is diminished. For $\theta_1$ above $2.0^\circ$, the spectrum is close to a simple Dirac cone with a linear slope and a point of vanishing density of states, despite the generic presence of a gap for spectra of the local Hamiltonians. Mini-band structures develop as $\theta_1$ is decreased below $2.0^\circ$ until even a true gap emerges in the spectrum below $1.8^\circ$. We note that $C_{2 z} \mathcal{T}$ does not contradict the opening of a gap as the pair of original Dirac cones is not protected (only a single isolated Dirac is).  $C_{2 z} \mathcal{T}$ however protects a Dirac point at $\bGM$.



\begin{figure}[!ht]
    \centering
    \includegraphics[width=0.9\linewidth]{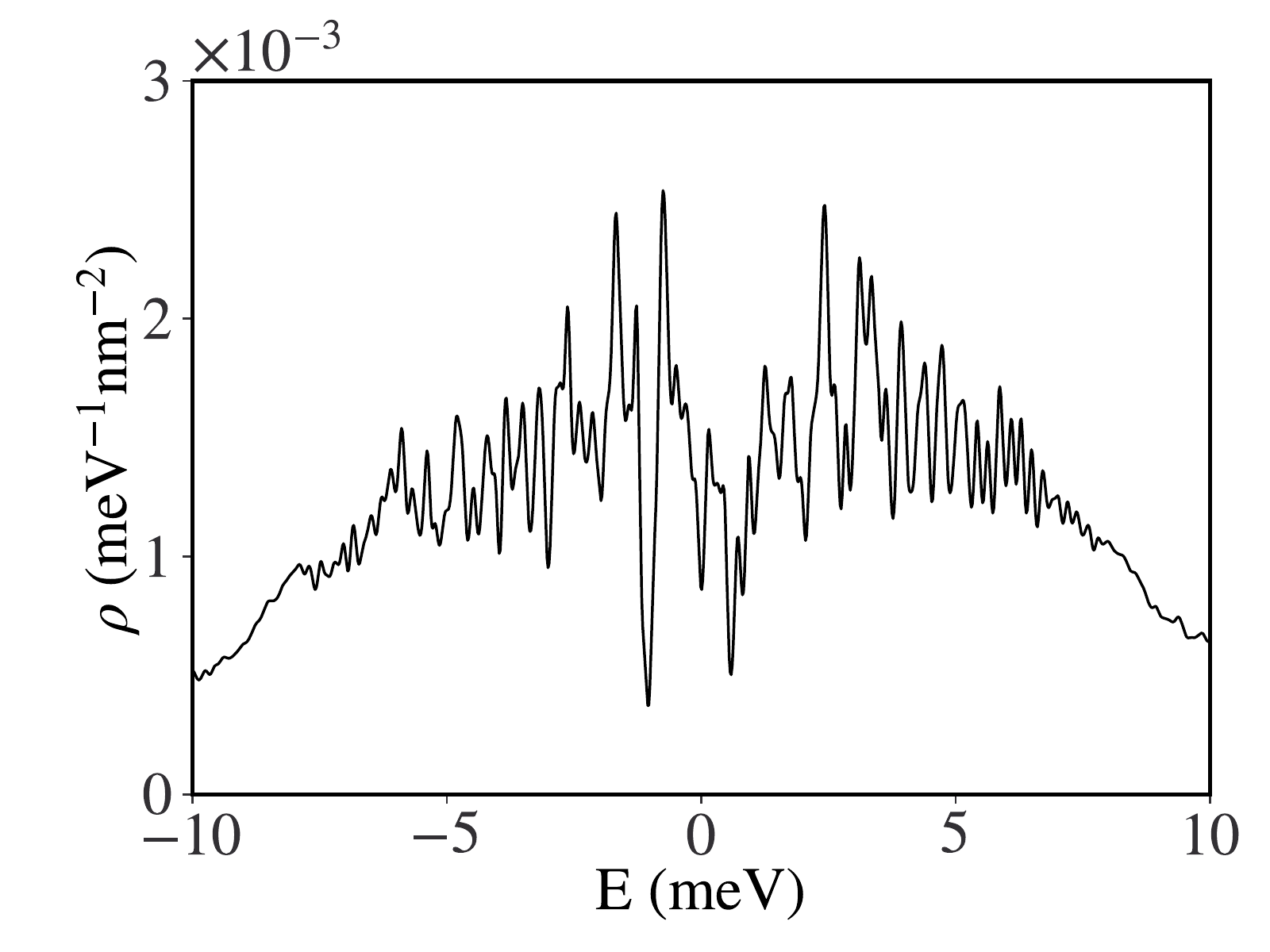}
    \caption{Density of states of the effective model at $\bGM$ with $\theta_1/\theta_2 = 1.3^\circ/2.6^\circ$.}
    \label{fig:dos unequal non-conv gamma}
\end{figure}
\begin{figure}[!ht]
    \centering
    \includegraphics[width=0.9\linewidth]{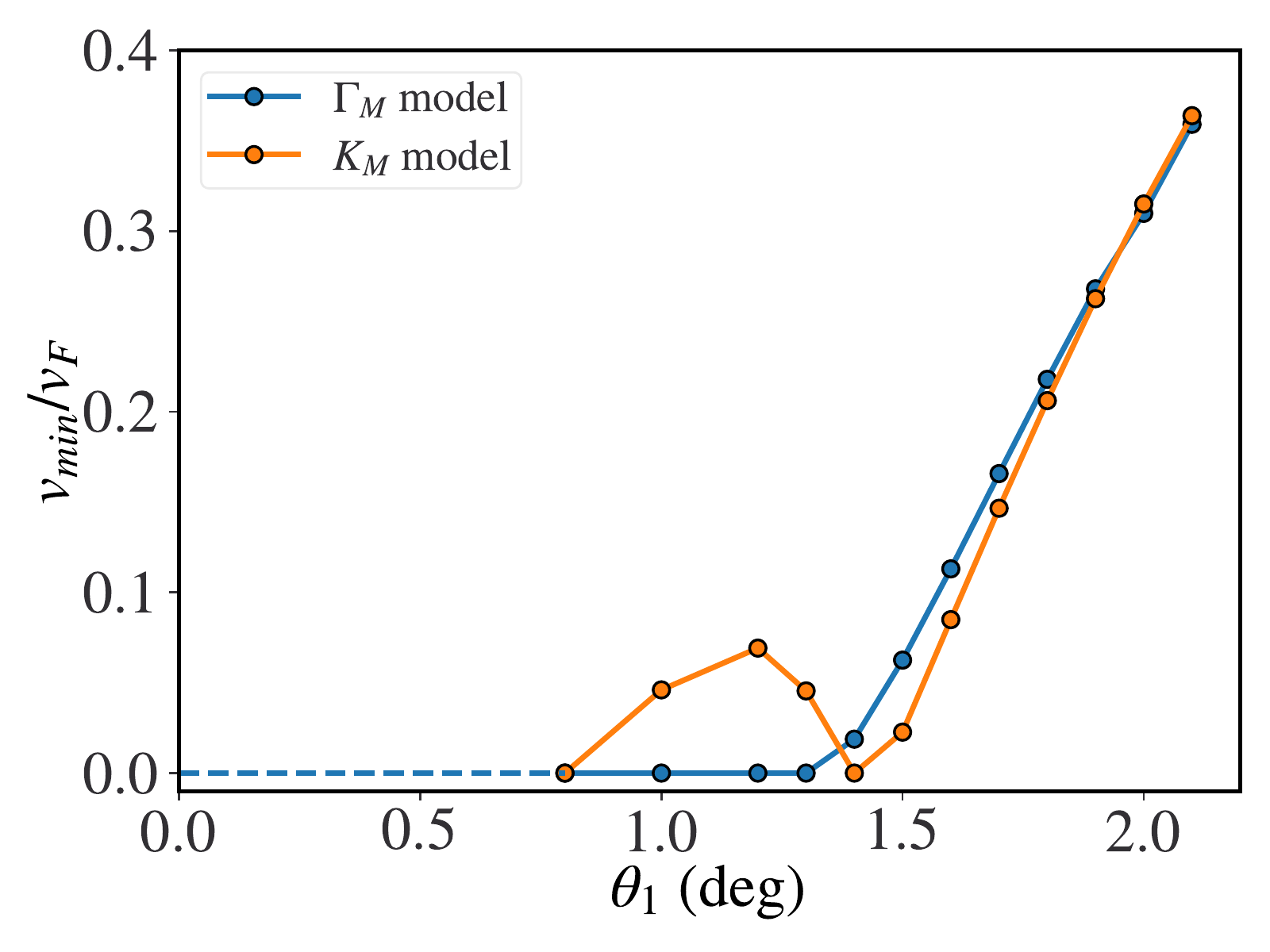}
    \caption{The minimum renormalized velocity components of the ${\bf \Gamma}_M$ and the ${\bf K}_M$ model as a function of $\theta_1 = \frac{1}{2}\theta_2$. The procedure for extracting the minimal velocity at the $K_M$ point with four states is detailed in Appendix~\ref{app:velocity}.
    \label{fig:vmin vs theta_1}}
\end{figure}
\begin{figure}[!ht]
    \centering
    \includegraphics[width=0.9\linewidth]{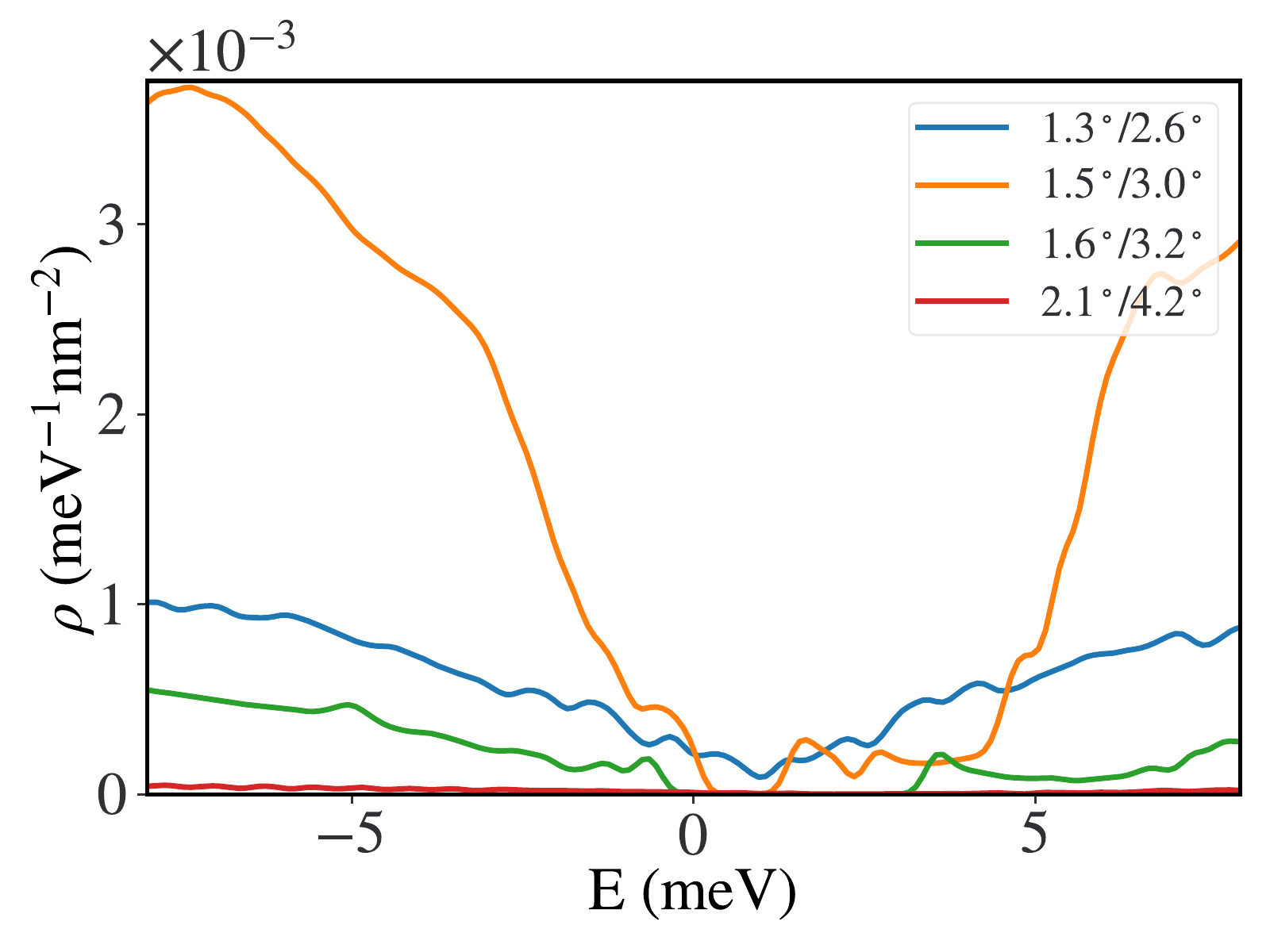}
    \caption{The $\bKM$ effective model density of states with different twist angles. Small Dirac cone velocity at $1.5^\circ/3.0^\circ$ gives rise to outstandingly high density of states in the low-energy region.}
    \label{fig:K unequal dos angles}
\end{figure}

Similarly to the case $\theta_1 \simeq \theta_2$, there are values of the twist angles where convergence with the number of momenta fails and they are correlated with a vanishing component of the Dirac cone velocity, see Appendix~\ref{app:velocity}. At ${\bf \Gamma}_M$, the absence of convergence starts below  $1.4^\circ$ and extends below as seen in Fig.~\ref{fig:vmin vs theta_1} where the minimal velocity is shown. The sign change of velocities is also associated with a strong enhancement of the density of states and a smearing of the Dirac dip as illustrated in Fig.~\ref{fig:dos unequal non-conv gamma}. At $\bKM$, the  minimal velocity also vanishes around $1.4^\circ$, see Fig.~\ref{fig:vmin vs theta_1}, but there is a revival for angles below. Interestingly, the revival of the velocity leads to a smaller and Dirac-cone like density of states, as shown in Fig.~\ref{fig:K unequal dos angles}. Hence, the total density of states exhibits a marked enhancement for angles around $1.4^\circ$, where the experiment of Ref.~\cite{Zhang2021Correlated} has been carried out, signaling a magic angle for TTG.

\begin{figure}[!ht]
    \centering
    \includegraphics[width=0.9\linewidth]{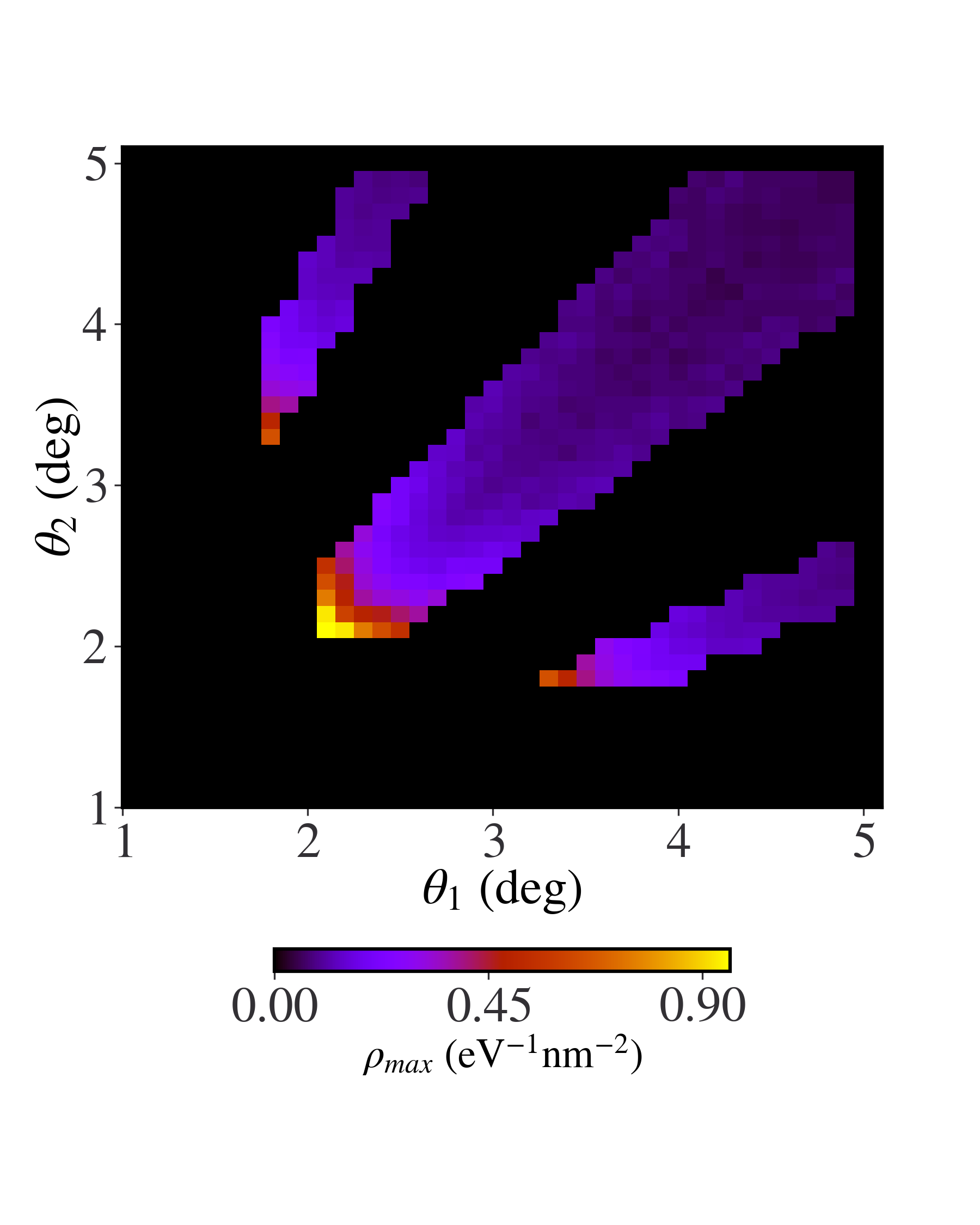}
    \caption{Maximum density of states found between $-30$ and $30$ meV calculated for the $(\theta_1, \theta_2)$ pairs satisfying the decoupling condition  $|\delta \bq|/|\bar{\bq}| \leq 0.2$.} 
    \label{fig:rho_max_angles}
\end{figure}

The generalization paves the way to the computation of many more pairs of $(\theta_1, \theta_2)$, but not arbitrary values of $\theta_1$ and $\theta_2$ because the decoupling condition proposed by Eq.~\eqref{moire_of_moire} is not always satisfied at arbitrary twist angles, which goes beyond the scope of our approach. Here we evaluate a series of maximum density of states at certain pairs of $(\theta_1, \theta_2)$ satisfying $\theta_1/\theta_2 \simeq 1/2, 1 ~\text{and}~ 2$. The energy range is limited to between $-30$ and $30$ meV.

The results of the numerical evaluations are shown in  Fig.~\ref{fig:rho_max_angles}. We set the criteria for the validity of decoupling \mr and \mom scales to be $|\delta \bq|/|\bar{\bq}| \leq 0.2$, so that the \mom unit cell area is at least $25$ times as larges as the \mr unit cell. The computation is performed for angles above 2 degrees for nearly equal angles, and for $\min(\theta_1, \theta_2) \geq 1.8^\circ$ if $\theta_1 / \theta_2 \simeq 1/2 ~\text{or}~ 2$. For smaller angles, the aforementioned zero-velocity singularity appears and our theory does not apply. Points in the $(\theta_1, \theta_2)$ plane that do not satisfy the above decoupling condition are not sampled .

\section{Conclusion}
\label{sec:conclusions}

We developed an analytical theory to describe the low-energy band spectrum of twisted trilayer graphene at the supermoir\'e scale. Our method projects the continuum Bistritzer-MacDonald model over local approximate eigenstates and obtain an effective theory which integrates out the details at the moir\'e scale and describe states that spread over the entire supermoir\'e lattice.

The effective model features a Dirac cone with a space-varying velocity coupled by a gauge field originating from the non-abelian Berry connection of the chosen local states. More precisely, out of the original three layers of graphene, we identified six low-energy regions, each described by a different effective model consisting of a (slightly gapped) Dirac cone equipped with a pseudo vector potential. The six regions correspond to the $\bKM/\bKM'$ valleys of graphene, each valley hosting three low-energy points at the high-symmetry positions $\bGM$, $\bKM$ and $\bKM'$ in the moir\'e Brillouin zone when the twist angles $\theta_1$, $\theta_2$ are almost equal.
The symmetries of the original continuum TTG model, in particular  $C_{3z}$  and $C_{2z}\mathcal{T}$, are locally broken but restored in the effective model. The model however does not possess particle-hole symmetry which shifts the Dirac cones at non-zero energy~\cite{cao2020tunable}.




In agreement with Ref.~\cite{PhysRevLett.123.026402} where an approximate solution to the model has been given, we observed that the spectrum of the effective model consists of fully connected bands and we prove that this feature is protected by the $C_{2z}\mathcal{T}$ symmetry of the three-layer graphene structure.
The diagonalization of the effective model showed that the density of states is very different from the incoherent addition of locally computed densities of states. It emphasizes the strong role played by the non-abelian gauge potential in reshaping the band structure to form mini-bands, especially when the typical velocity becomes very small. The weight redistribution is important already for twist angles around $2.4^\circ$. At angles below $2^\circ$, we find that very small velocities smear out the Dirac cone dip in the density of states and result in an accumulation of low-energy states.

For such small velocities, the effective model develops a singularity indicating that low-energy wavefunctions change rapidly over the moir\'e scale in this regime. This signals the importance of short-range moir\'e physics in twisted trilayer graphene at low twist angles. We investigated the two cases $\theta_1 \simeq \theta_2$ and $\theta_1/\theta_2 \simeq  1/2$, where a similar increase in the density of states was obtained, but our method would apply equally to $\theta_1/\theta_2$ close to any rational number.

The effective theory detailed in this work is in fact very general and can be extended to any band models with a spatial scale separation. This would include more general multilayer structures where the presence of two non-commensurate modulations gives rise to a supermoir\'e structure. This is particularly relevant to twisted bilayer graphene with a slightly misalign substrate such as hBN~\cite{shi2021moire}. Here the supermoir\'e structure gives rise to a spatial map of
distinct states, including correlated insulating states with various Chern numbers, semimetal states, and valley-unpolarized
states~\cite{Grover2022}. Coherent band structure effects at this long-wavelength might reveal interesting phenomena provided that the electronic mean-free path due to impurity scattering is longer than the supermoir\'e wavelength.

\begin{acknowledgments}
We thank Felix von Oppen, Andrei Bernevig, Nicolas Regnault, Jen Cano, Jed Pixley and Ziyan Zhu for valuable discussions. The Flatiron Institute is a division of the Simons Foundation. We acknowledge support by the French National Research Agency (project TWISTGRAPH, ANR-21-CE47-0018).

\end{acknowledgments}

\appendix

\section{Extracting the velocities at the Dirac point}
\label{app:velocity}

We first discuss a general anisotropic Dirac cone Hamiltonian: 
\begin{equation}
h(\bk) = \sum_{i,j} \tau_i d_{ij} k_j ~,
\end{equation}
with $i \in {x,y,z}$ and $j \in {x,y}$. $\tau_i$ are the Pauli matrices defined on the indices of the two states chosen to build the effective model. Then $d$ is a $3\times 2$ matrix and admits a singular-value decomposition (SVD):
\begin{equation}
d = U^T S V,
\end{equation}
with $U$ and $V$ being orthogonal matrices and $S$ a $3\times2$ diagonal matrix. The diagonal elements of $S$ matrix contains two singular values of $d$, noted $v_1$ and $v_2$. 
So we have:
\begin{equation}
\begin{split}
    & U_{ij} U_{kj} = U_{ji} U_{jk} = \delta_{ij} ~,\\
    & V_{ij} V_{kj} = V_{ji} V_{jk} = \delta_{ij} ~,\\
    & S_{ij} = v_i \delta_{ij} ~.
\end{split}
\end{equation}
We can show that the singular values are the two components of the velocity characterizing  the cone.
Then let us define: 
\begin{equation}
\tau'_i = \sum_{j \in \{x,y,z\}} U_{ij} \tau_j ~,
\end{equation}
and 
\begin{equation}
k'_i = \sum_{j \in \{x,y\}} V_{ij} k_j ~.
\end{equation}
The rotated matrices $\tau'_i$ also obeys the Clifford algebra of Pauli matrices
\begin{equation}
\tau'_i \tau'_j = i\varepsilon_{ijk}.\tau'_k + \delta_{ij}\mathbb{I}, \quad i,j,k \in \{1,2,3\} ~.
\end{equation}
Proof: 
Since with the Pauli matrices we have $\tau_i \tau_j = i\varepsilon_{ijk}.\tau_k + \delta_{ij}\mathbb{I}, \quad i,j,k \in \{x,y,z\} $, then (adopting the automatic contraction of the same indices)
\begin{equation}
\begin{split}
\tau'_i \tau'_j &= U_{il} U_{jm} \tau_l \tau_k = U_{il} U_{jm}(i\varepsilon_{lmp} \tau_p + \delta_{lm}\mathbb{I}) \\
& = i\varepsilon_{lmp} U_{il} U_{jm} \tau_p + U_{il} U_{jm} \delta_{lm}\mathbb{I} \\
& = i\varepsilon_{ijk} U_{kp} \tau_p + \delta_{ij}\mathbb{I} \\
& = i\varepsilon_{ijk} \tau'_k + \delta_{ij} \mathbb{I}
\end{split} ~.
\end{equation}
The Dirac Hamiltonian becomes:
\begin{equation}
h(\bk) \rightarrow h(\bk') =  v_1 \tau'_1 k'_1 + v_2 \tau'_2 k'_2 .
\end{equation}
In this rotated basis, $v_1$ and $v_2$ emerge as the two velocity components of the anisotropic Dirac cone.


When a high-symmetry point hosts four degenerate states, the $\bk \cdot {\bf p}$ method gives the low-energy Hamiltonian
\begin{equation}
    H_{\bk} = {\bm \Gamma} \cdot \bk = \Gamma_x k_x + \Gamma_y k_y
\end{equation}
with the two $4\times4$ matrices $\Gamma_x$, $\Gamma_y$. Four angular-dependent velocities can be defined from the spectrum eigenvalues as
\begin{equation}
    v_j (\theta) = \frac{{\rm eig}_j ({\bm \Gamma} \cdot \bk)}{|\bk|} \qquad j = 1,\ldots,4
\end{equation}
where $\theta$ is the polar angle of $\bk$. The minimal velocity is obtained by minimizing $v_j (\theta)$ over $j$ and $\theta$.

\begin{figure}[!ht]
    \centering
    \includegraphics[width=\linewidth]{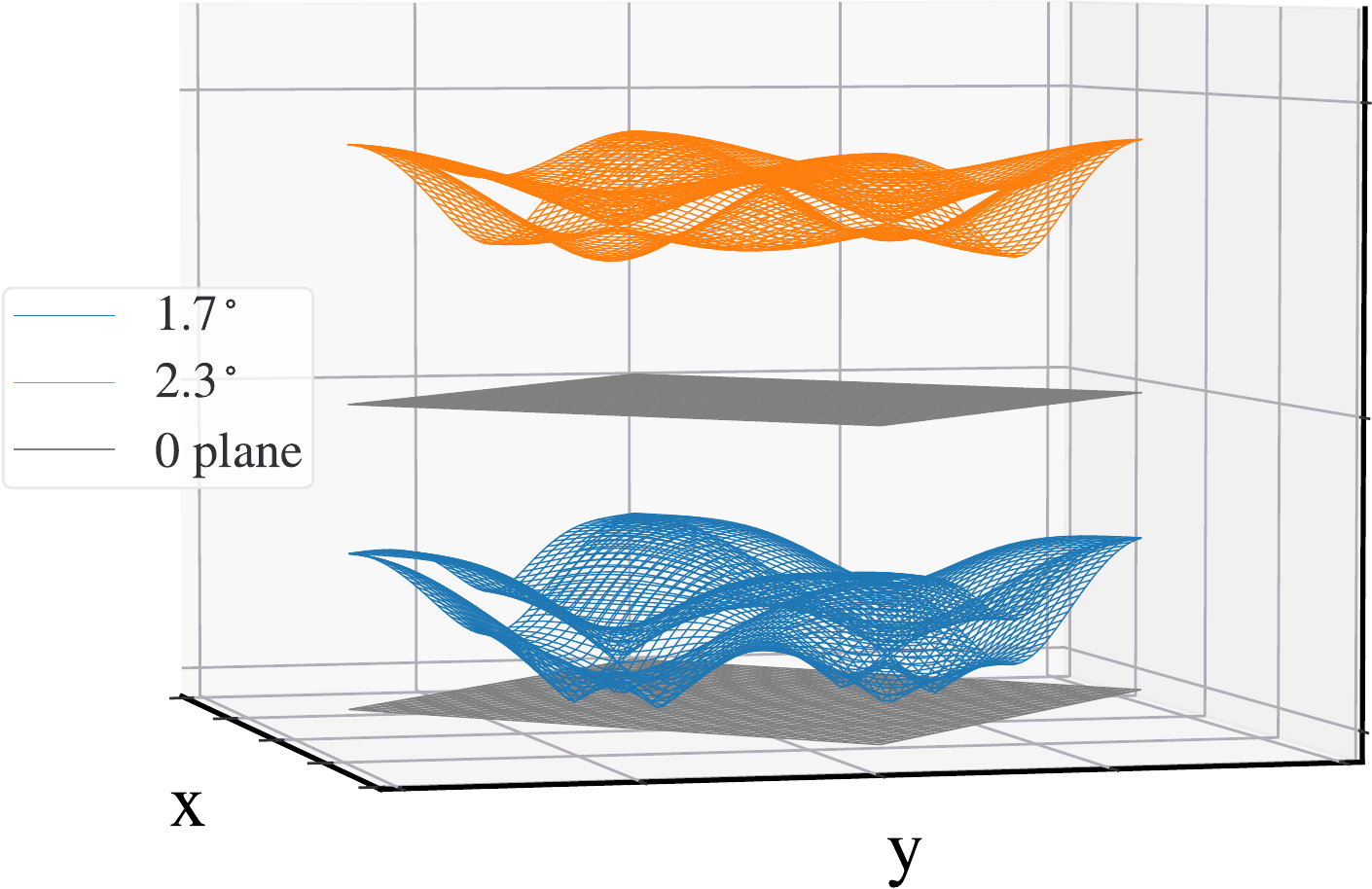}
    \caption{Variation of the velocity of the local Dirac cone across a \mom unit cell. The velocity profiles are shown for two cones at $\Gamma$ point for equal twist angles of $1.7^\circ$ and $2.3^\circ$. At small twist angles, one component of the velocity can drop to zero, causing a singularity in the wavefunction. At the AAA- and ABA-stacking points, the two components become equal and an isotropic Dirac cone is restored. }
    \label{fig:velocity in R}
\end{figure}

In Fig.~\ref{fig:velocity in R}, we show the velocity components of the Dirac cone measured at different locations of $\bR$, for equal twist angles $2.3^\circ$ and $1.7^\circ$. When the twist angle becomes small, the smaller component will reach zero value at certain coordinates, which become a singularity in the wavefunction that will be discussed in Appendix.~\ref{app:1Dtoy}.

\section{Symmetries and gauge-invariance of the effective model}
\label{app:symmetry}
\textit{Global and local symmetries.}~ The local Hamiltonians, Eqs.~\eqref{H_trilayer_staircase3},~\eqref{H_trilayer_staircase2_unequal}, break essentially all spatial symmetries - except at certain high-symmetry points - even though these symmetries are satisfied in the original model and restored in the effective model Eq.~\eqref{eff_hamil2}.

More precisely, consider a given symmetry represented by the operator  $\hat S$ and acting as 
$ \hat S H_K  \hat S^\dagger = H_K$ on the original Hamiltonian Eq.~\eqref{H_trilayer_staircase}. The symmetry does not leave the local Hamiltonian invariant but maps it to a different position $O_{[S]} \bR$,
\begin{equation}\label{eq:determin eff symmetry}
    \hat S H_M(\phi(\bR)) \hat S^\dag = H_M(\phi(\hat O_{[S]} \bR)),
\end{equation}
$\hat O_{[S]}$ being the symmetry operation on the $\bR$ coordinate.
$\hat S$ is in general \emph{not} a symmetry of the local Hamiltonian unless $\hat O_{[S]} \bR = \bR$. Yet it implies that the local spectrum is invariant upon sending $\bR$ to  $\hat O_{[S]} \bR$. With $H_M( \phi(\bR)) \ket{\psi_{\alpha \bk; \phi(\bR)}} = \mathcal{E}_{\alpha \bk}(\bR) \ket{\psi_{\alpha \bk; \phi(\bR)}}$, Eq.~\eqref{eq:determin eff symmetry} implies that $H_M(\phi(\hat O_{[S]} \bR)) \hat S \ket{\psi_{\alpha\bk; \phi(\bR)}} = \mathcal{E}_{\alpha \bk} (\bR) \hat S \ket{\psi_{\alpha \bk;\phi(\bR)}}$, meaning that $H_M(\phi(\bR))$ and $H_M(\phi(\hat O_{[S]}\bR))$ share the same spectrum. We can relate the corresponding eigenstates
\begin{equation}\label{eq:sym transform local states}
    \ket{\psi_{\alpha \bk; \phi(\hat O_{[S]} \bR)}} =  \sum_{\alpha'} \hat S \ket{\psi_{\alpha' \bk; \phi(\bR)}} W_{\alpha' \alpha}
\end{equation}
where the unitary matrix $W$ depends on the choice of gauge both at the positions $\bR$ and  $\hat O_{[S]} \bR$.
In the following we shall omit the subscript $\bk$ for simplicity since the local states are taken from different bands at the same $\bk$ point in the \mr BZ. 

The effective low-energy Hamiltonian~\eqref{eff_hamil2} acts on the $\bR$ coordinate with matrix elements evaluated at $\bR$. Operating with the rotation $\hat O_{[S]} \bR$ on $H_{\rm eff}$, we denote ${\cal S}$ the induced symmetry operator, with 
\begin{equation}
    {\cal S} H_\text{eff} {\cal S}^\dagger \equiv \mathcal{E}(\hat O_{[S]} \bR) +\frac{1}{2}\left\{\bm \Gamma(\hat O_{[S]} \bR), \hat O^{-1}_{[S]} \hat{\bk} \right\}  + A(\hat O_{[S]} \bR) ,
\end{equation}
where the matrix elements in Eq.~\eqref{gamma},~\eqref{nonabelian} are computed at the rotated position $\hat O_{[S]} \bR$. Using Eq.~\eqref{eq:sym transform local states}, we find for the first term
\begin{equation}\label{trans1}
\begin{split}
\mathcal{\bm E}_{\alpha \beta} (\hat O_{[S]} \bR) &= \mel{\psi_{\alpha;\hat O_{[S]} \bR}}{H_M(\hat O_{[S]} \bR)}{\psi_{\beta; \hat O_{[S]} \bR}} \\
&= \sum_{\alpha' \beta'} W^\dag_{\alpha \alpha'}\mel{\psi_{\alpha;\bR}}{H_M(\bR)}{\psi_{\beta; \bR}} W_{\beta' \beta} \\
&= \sum_{\alpha' \beta'}  W^\dag_{\alpha \alpha'} \mathcal{E}(\bR)_{\alpha' \beta'} W_{\beta' \beta} .
\end{split} 
\end{equation}
The last two terms do not transform separately with $W$ but their combination does, 
\begin{equation}\label{trans2}
\begin{split}
    \frac{1}{2}\left\{\bm \Gamma(\hat O_{[S]} \bR), \hat O^{-1}_{[S]} \hat{\bk} \right\}  + A(\hat O_{[S]} \bR) \\
    = W^\dag \left( \frac{1}{2}\left\{\bm \Gamma(\bR), \hat{\bk} \right\}  + A(\bR) \right) W .
\end{split}
\end{equation}
Combining Eqs.~\eqref{trans1} and~\eqref{trans2}, we find that the effective model transforms as
\begin{equation}\label{eq:eff symmetry}
    {\cal S} H_\text{eff} {\cal S}^\dagger = W^\dag H_\text{eff}(\bR) W ,
\end{equation}
where $W$ depends on the gauge choice of local states. Eq.~\eqref{eq:eff symmetry} unambiguously proves that the spectrum of the effective low-energy model Eq.~\eqref{eff_hamil2} is invariant under the symmetry. We now detail the different symmetries.

\emph{$C_{3z}$  and $ C_{2z}\mathcal{T}$ symmetry}. These two symmetries operate as
\begin{equation}
\begin{split}
    \hat C_{3z} H_M(\phi(\bR)) \hat C^\dag_{3z} = H_M(\phi(\hat R_\frac{2\pi}{3} \bR)) , \\ (\hat C_{2z}\mathcal{T})  H_M(\phi(\bR)) (\hat C_{2z}\mathcal{T})^\dag = H_M(\phi(-\bR)) .
\end{split}
\end{equation}
with
\begin{equation}
\begin{split}
    \hat C_{3z} = e^{2i\pi\sigma_z/3}  \delta_{\hat R_{2 \pi/3} \br,\br'}  \\
    \hat C_{2z}\mathcal{T} = \sigma_x K \delta_{-\br,\br'}  ~ ,
\end{split} 
\end{equation}
$\hat R_\theta$ rotates a coordinate by the angle $\theta$ around the $z$ axis, $K$ is the complex conjugation operator.

\emph{$C_{2x}$ symmetry.}~ $C_{2x}$ is a symmetry only for $\theta_1 \simeq \theta_2$, and acts as
\begin{equation}
    \hat C_{2x} H_M(\phi(\bR)) \hat C^\dag_{2x} = H_M(\phi(\hat M_1 \bR)).
\end{equation}
$\hat M_1$ rotates $\bR$ by $180^\circ$ around the $x$ axis and
\begin{equation}
    \hat C_{2x} = \sigma_x \begin{pmatrix}
        & & 1 \\
        & 1 & \\
        1 & & 
    \end{pmatrix}_\text{layer}  \delta_{\hat M_1 \br,\br'}   ~
\end{equation}

\emph{Translational symmetry.}~
The symmetry by translation is relevant for the periodicity of the effective model. One can show that for arbitrary ratio of $\theta_1/\theta_2$, there always exists a translation operator $e^{i\hat{\bk}\br_0}$ such that 
\begin{equation}
    e^{i\hat{\bk}\br_0} H_M(\phi(\bR)) e^{-i\hat{\bk}\br_0} = H_M(\phi(\bR + \bm a^{MM}_0)) ,
\end{equation}
where $\hat{\bk} = -i \bm \nabla$ is an operator defined on the \emph{\mr} coordinate $\bf r$. 
The values of $\br_0$ and $\bm a^{MM}_0$ depend on the ratio $\theta_1/\theta_2$. The set of admissible values for  $\bm a^{MM}_0$ defines the \mom Bravais lattice.

\emph{Gauge invariance.}~ A change of gauge is a special case where $\hat S$ is a pure phase in Eq.~\eqref{eq:sym transform local states}, while $W$ is determined by the two different gauge choices. In this case, Eq.~\eqref{eq:eff symmetry} directly proves the  \emph{gauge-invariance} of the effective model.

\section{Proof of the full connectivity of the trilayer graphene band structure}
\label{app:fullconnection}

In this appendix, we give a detailed proof of why the spectrum of twisted trilayer graphene is fully connected as a consequence of the $C_{2z}\mathcal{T}$ symmetry (see also Refs.~\cite{ahn2019failure,PhysRevLett.123.026402}). We shall detail three arguments which have a common topological origin.

The first argument is in essence analogous to the proof of the absence of gap opening on the surface of a $\mathbb{Z}_2$ (class AII) 3D topological insulator by a decorating moir\'e potential~\cite{cano2021moire}. The Dirac cone on the surface cannot be gapped out by a time-reversal preserving perturbation, including a moir\'e potential, as the non-trivial bulk topology requires edge states connecting the conduction and valence bands. The spectrum is thus fully connected despite the mini-band folding~\cite{cano2021moire}. A single Dirac cone, or an {\it odd} number of Dirac cones, can also be protected on the surface of a 3D topological crystalline insulator (TCI) with $C_{2z}\mathcal{T}$ symmetry~\cite{fang2015new}, the $z$-axis being normal to the surface. If we now imagine that TTG arises on the surface of a fictitious $C_{2z}\mathcal{T}$-preserving CTI or, more precisely, that each monolayer Dirac cone (in valley $K$) is a surface state of the CTI, then the same argument of bulk topology protects a fully-connected spectrum since interlayer electron tunneling respects the $C_{2z}\mathcal{T}$ symmetry.



The second argument is more direct. A finite and isolated set of bands, in a model satisfying $C_{2z}\mathcal{T}$, is characterized by an integer topological invariant~\cite{ahn2019failure} known as the second Stiefel Whitney class $w_2$. On the other hand for two bands, $w_2$ coincides modulo two with another topological index, the Euler class $e_2$. The latter is given by $e_2=\oint_{\mathcal C}{d\mathbf k}\cdot {\mathbf A}_{+,-}({\mathbf k})/2\pi$ where the line integral encloses the Dirac cone, ${\mathbf A}_{+,-}({\mathbf k})=i\mel{u_{\mathbf k +}}{\nabla_{\mathbf k}}{u_{\mathbf k -}}=(-k_y,k_x)/(2k^2)$ and $\ket{u_{\mathbf k \pm}}=\left[1,\pm (k_x+ik_y)/k\right]^T/\sqrt{2}$ are the eigenstates of the Dirac Hamiltonian. Performing straightforward calculations we readily find:
\begin{equation}
    e_2 = \frac 1 2.
\end{equation}
Turning on the hopping between layers which preserves $C_{2z}\mathcal{T}$, there is no way a combination of three Dirac cones with Euler class $\pm 1/2$ can transform into an integer $w_2$. Therefore, by contradiction, there is no finite set of isolated bands and the spectrum must be fully connected.

The third argument is analogous to the second but proceeds more constructively. We assume again a finite set of $N$ isolated bands with $C_{2z}\mathcal{T}$ symmetry. 
We first prove that the $N$ bands exhibits an even number of Dirac crossings within them.
Because of the  $C_{2z}\mathcal{T}$ symmetry, the local Berry curvature, evaluated for any band, is vanishing except at band crossing points~\cite{hejazi2019,song2019}. Computing the Wilson loop for a given band, this implies that the corresponding Berry phase is constant as one moves the loop across the Brillouin zone, except at Dirac point crossings - with the band above or the band below - where the Berry phase jumps by $\pm \pi$. As the Wilson loop Berry phase must return to its original value after sweeping across the whole Brillouin zone, it proves that the number of Dirac crossing from one band to its neighbors must be an even integer. In particular, the first band (with the lowest energy) has an even number of Dirac crossings with the second. It directly implies the same between the second and the third, and by recurrence between any $i$ and $i+1$. As a result, the total number of Dirac band crossings must be even. As Dirac cones can only fuse or be created by pairs under $C_{2z}\mathcal{T}$, we find that the $N$ bands cannot be continuously connected to the original three Dirac cones in valley $K$. By contradiction, it proves the fully connected spectrum without any gap between bands.

\section{Manual to the numerical solution of the effective model}\label{app:manual}

We briefly explain the numerical procedure for solving the effective model given in Eq.~\eqref{eff_hamil2}. As the different coefficients in Eq.~\eqref{eff_hamil2} - such as ${\bm \Gamma} (\bR) $ or $A(\bR)$ - are obtained by diagonalizing independent local Hamiltonians from Eq.~\eqref{H_trilayer_staircase3}, a gauge-fixing method is required to work with smoothly varying functions of $\bR$. This choice of continuous functions for the coefficients is a prerequisite for the numerical solution. It is readily done in momentum space, using the periodicity of all coefficients on the \mom lattice.

\textit{Gauge fixing.} We first discuss the procedure at $\bGM$ where we have two degenerate zero-energy solutions to Eq.~\eqref{zero-energy} and arbitrary linear combinations can be chosen at each position $\bR$. We first introduce the two basis states of the local Hamiltonians, $|1\rangle = \ket{\bm Q = \bm 0, l = 0, A}$ and $|2\rangle = \ket{\bm Q = \bm 0, l = 0, B}$, where $l = -1, 0, 1$ denotes the layer index: bottom, middle and top, $A/B$ the sublattice index. These two states $|g\rangle$ ($g=1,2$) are not eigenstates of the local Hamiltonians, they are spatially homogeneous ($\bm Q=0$) over the moiré lattice. Collecting the numerical solutions of Eq.~\eqref{zero-energy}, we rotate them to a set of new states
\begin{equation}
  v_{\Gamma,\bR,\beta} (\br) = \sum_{\beta'=1,2} G_{\beta,\beta'} (\bR)
  u_{\Gamma,\bR,\beta} (\br) 
\end{equation}
with a set of unitary matrix $G (\bR)$ chosen such as to satisfy the constraints
\begin{equation}
 \mel{v_{\Gamma,\bR,\beta}}{\sigma_z}{v_{\Gamma,\bR,\beta}} \propto \delta_{\beta,\beta'}  ,
\end{equation}
and
\begin{equation}
   \braket{g}{v_{\Gamma,\bR,\beta}} \in \mathbb{R}^+,
\end{equation}
at each position $\bR$ and using the two reference states $g=1,2$. Computing the matrix coefficients $\mathcal{E} (\bR)$, ${\bm \Gamma} (\bR) $ and $A(\bR)$ with the rotated zero-energy states $v_{\Gamma,\bR,\beta} (\br)$, we obtain smooth and periodic functions. We have used the same procedure at $\bKM$ in Sec.~\ref{sec:bands_equal} but with the reference states $\ket{\bm k = \bm 0, l = 1, A}$ and $\ket{\bm k = \bm 0, l = 1, B}$.

\textit{Fourier transform.}~ The gauge-fixing condition yields an effective Hamiltonian varying smoothly with the position $\bR$. In addition, its periodicity over the \mom unit cell, derived in Sec.~\ref{subsec:symmetries}, implies a Bloch decomposition
\begin{equation}
    \begin{pmatrix}
f_1 (\bR) \\ f_2 (\bR)
\end{pmatrix} = e^{i \bk \cdot \bR} \begin{pmatrix}
w_1 (\bR) \\ w_2 (\bR)
\end{pmatrix},
\end{equation}
where the periodic functions $w_{1,2} (\bR)$ can be expanded into a (Fourier) discrete set of momenta generated by the reciprocal vectors ${\bf b}_1^{MM} = \delta \bq_2 - \delta \bq_1$ and ${\bf b}_2^{MM} = \delta \bq_3 - \delta \bq_1$ of the \mom lattice. The following numerical procedure is then the same as for the spectrum of twisted bilayer graphene~\cite{bistritzer2011moire} or of the local Hamiltonians Eq.~\eqref{H_trilayer_staircase3}: the Hamiltonian $H_{\rm eff}$ couples the different Fourier components of $w_{1,2} (\bR)$, with an exponential convergence of the spectrum with the number of kept momenta $N_c$. Eventually, the effective Hamiltonian $H_{\rm eff}$ in Eq.~\eqref{eff_hamil} and Eq.~\eqref{eff_hamil2} takes the form of an $2 N_c \times 2 N_c$ matrix in this momentum lattice for each value of $\bk$ spanning the \mom Brillouin zone. 

\section{Protection of Dirac cone at \texorpdfstring{$\bGM$}{Gamma} of local \mr models}
\label{app:ph_symmetry}

For $\theta_1 \simeq \theta_2$, the local Hamiltonian Eq.~\eqref{H_trilayer_staircase3} satisfies a particle-hole symmetry which pins a Dirac cone at zero energy at the $\bGM$ point regardless of the value of $\bR$ ({\it i.e.} $\phi$), see Fig.~\ref{fig:local spectra}. However and in contrast with the spatial symmetries, the starting continuum model Eq.~\eqref{H_trilayer_staircase} and the effective low-energy model Eq.~\eqref{eff_hamil2} do not exhibit a particle-hole symmetry.

We discuss here the protection of the Dirac cone by particle-hole symmetry. The mechanism itself is interesting because it is unconventional. Its consequence is also significant as it cancels at $\bGM$ the energy term ${\cal E} (\bR)$ in Eq.~\eqref{eff_hamil}.

\textbf{Equal twist angles.}~ For $\theta_1 \simeq \theta_2$, the particle-hole symmetry governs the local Hamiltonian
\begin{equation}\label{Psymm}
    \mathcal{P} H_M(\phi) \mathcal{P}^\dag = -H_M(\phi)
\end{equation}
with
\begin{equation}\label{matrixP}
    \mathcal{P} =  \mathcal{P}^\dagger = \begin{pmatrix}
        & & 1 \\
        & -1 & \\
        1 & &
    \end{pmatrix}_\text{layer} \delta_{-\br,\br'},
\end{equation}
and $\mathcal{P}$ is diagonal in the sublattice index. The representation of $\mathcal{P}$ within the set of eigenstates of $H_M(\phi)$ is best discussed for $\alpha=0$, {\it i.e.} when the tunnel coupling between layers is turned off. In this case, the model (see Fig.~\ref{fig:uneuqal coupling}) exhibits pairs of zero-energy states, at $\bGM$ (middle layer), $\bKM$ (top layer) and $\bKM'$ (bottom layer), from which Dirac cones emerge. The way states transform under ${\cal P}$ is read off Eq.~\eqref{matrixP}, or
\begin{equation}\label{eq:PHS transfrom}
\begin{split}
    & \mathcal{P} \ket{\bGM, a} = - \ket{\bGM, a}, \\
    & \mathcal{P} \ket{\bKM, a} = \ket{\bKM', a}, \\
    & \mathcal{P} \ket{\bKM', a} = \ket{\bKM, a}, \\
\end{split}
\end{equation}
where $a=1,2$ labels the two states in each pair. From Eq.~\eqref{eq:PHS transfrom}, it is easy to compute the character of ${\cal P}$ within the complete Hilbert space. Since pairs of opposite momentum $\bk$ and energy $E$ transform onto each other by ${\cal P}$, or 
\begin{equation}\label{symstate}
    \mathcal{P} \ket{\bk, E} = \ket{-\bk, -E}
\end{equation}
as implied by Eq.~\eqref{Psymm}, ${\cal P}$ projected onto each pair has a vanishing trace. The same is true at $\bKM$ and $\bKM'$. The only non-zero contribution to the trace of ${\cal P}$ thus comes from  $\bGM$ so that overall ${\rm Tr} {\cal P} = -2$. 

At $\alpha \ne 0$, Eq.~\eqref{symstate} still holds such that only zero-energy states are relevant for ${\rm Tr} {\cal P}$. ${\cal P}$ is an involutory matrix, ${\cal P}^2 = 1$, with eigenvalues $\pm 1$ and its trace is thus an integer. By continuity, this argument shows that ${\rm Tr} {\cal P} = -2$ for arbitrary $\alpha$. If we assume that the two states at $\bGM$ leave the zero-energy subspace, and they can only leave as particle-hole partners, then suddenly ${\rm Tr} {\cal P} = 0$ in contradiction with continuity. This completes the proof that ${\cal P}$ protects the Dirac point at $\bGM$.

\textbf{Unequal twist angles.}~ In contrast with the previous case, there is no general particle-hole symmetry for unequal twist angles especially for $\theta_1 / \theta_2$ close to $1/2$. There is however another particle-hole symmetry which protects a Dirac cone only along three lines in the \mom unit cell: the $y$ axis (at $x=0$) and its symmetric partners by $C_{3 z}$ and $C_{3 z}^{-1}$. The symmetry operator reads
\begin{equation}
    \mathcal{P}_1 = \sigma_x \begin{pmatrix}
        1 & & \\
        & -1 & \\
        & & 1
    \end{pmatrix}_\text{layer} \delta_{\hat M_y \br,\br'},
\end{equation}
with the mirror symmetry $\hat M_y (x,y) = (-x,y)$, effectively exchanging $\phi_2$ and $\phi_3$, and the local Hamiltonian transforms as
\begin{equation}
    \mathcal{P}_1 H_M(\phi) \mathcal{P}^\dag_1 = - H_M(\hat M_y \phi) ~,
\end{equation} 
 $\mathcal{P}_1 $ acts as a true particle-hole symmetry for $\phi_2 = \phi_3$ (such that $\phi = \hat M_y \phi$).
Resuming the above arguments, we find that $\mathcal{P}_1 $  pins a Dirac cone at zero energy and it must remain at $\bGM$ due to the additional $C_{3z}$ symmetry. For $\phi_2 - \phi_1 = \phi_3 - \phi_2 = \pm 2\pi/3$,  $\mathcal{P}_1 $ combined with $C_{3z}$  protects a Dirac cone at $\bGM$ along the two $C_{3 z}$ partners of the $y$-axis.

\section{Vanishing Dirac velocity}
\label{app:1Dtoy}

We discuss here a 1D Dirac model with a space-dependent velocity and show that a sign change of the velocity induces a singularity. The same argument is detailed in Refs.~\cite{takahashi2011gapless, buccheri2022transport, christophe2021artificial} in the context of a singular Klein tunneling and an associated anomaly.
Consider the 1D hermitian Dirac Hamiltonian 
\begin{equation}
    h(x) = \sqrt{v(x)} (-i\partial_x) \sqrt{v(x)},
\end{equation}
where $v(x)$ is an arbitrary smooth and positive continuous function. Eigenstates with the energy $\varepsilon$ have an analytical expression
\begin{equation}
    \varphi(x) = \frac{\mathcal{N}}{\sqrt{v(x)}}  e^{i \int^x \dd x' \frac{\varepsilon}{v(x')}},
\end{equation}
where $\mathcal{N}$ is a normalizing factor. This wavefunction is obviously singular whenever the velocity vanishes. The corresponding density of states then exhibits a logarithmic singularity.



\bibliography{references}
\end{document}